\documentclass[11pt,draftcls,onecolumn]{IEEEtran} 

\usepackage{cite}      

\usepackage{graphicx}  

\usepackage{psfrag}    

\usepackage{subfigure} 

\usepackage{url}       

\usepackage{stfloats}  
\usepackage{hyperref}

\usepackage{amsmath}   
\usepackage{amsmath,amssymb,graphicx,url,color,bbm, amsthm,amsfonts,dsfont,array,psfrag,stmaryrd,wasysym,algorithmic,caption}

\usepackage[ruled,vlined]{algorithm2e}

\DeclareMathOperator{\Id}{Id}

\DeclareMathOperator{\prox}{prox}

\DeclareMathOperator*{\argmin}{{arg\,min}\ }

\DeclareMathOperator{\sparsPrior}{\mathcal P}

\makeatletter
\DeclareRobustCommand\onedot{\futurelet\@let@token\@onedot}
\def\@onedot{\ifx\@let@token.\else.\null\fi\xspace}

\def\eg{\emph{e.g}\onedot} 
\def\ie{\emph{i.e}\onedot}

\def\etal{\emph{et al}\onedot}

\def\prox{\textit{prox}}

\def\Psia{\Psi_{\text{2D}}}

\def\mhat{\widehat m}

\def\tcal{\mathcal T}
\def\acal{\mathcal A}
\def\atilde{\widetilde A}

\def\hbf{\mathbf H}

\def\usm[#1]{\atilde_{#1}}

\def\up{\mathcal U}
\def\lo{\mathcal L}
\def\Thetabf{\mathbf{\Theta}}
\def\Thetan{\Theta}
\def\Psibf{\mathbf{\Psi}}
\def\Psin{\Psi}
\def\u{\mathbb{I}}
\def\k{k'}
\def\D{\mathbf D}
\def\sbf{\mathbf S}

\def\O{\mathcal O}

\def\na{n_{1}}
\def\nb{n_{2}}

\def\opAPsi{L}

\def\residdecorMtx{\mathbf{r}}

\newcommand{\defeq}{\triangleq}

\newcommand{\Rbb}{\mathbb{R}}
\newcommand{\Nbb}{\mathbb{N}}

\newcommand{\offdiag}{\text{Off diag}}

\newcommand{\mixMatrix}{\hbf}

\def\tabvspace{0.1cm}

\newcommand\fastSSIHTfinal{\texttt{SS-IHT}}

\newcommand\solveBPDNone{\texttt{BPDN}}
\newcommand\solveTVDN{\texttt{TVDN}}

\newcommand\fastSSstIHTfinalProj{\texttt{SS-IHT-decorr}}

\newcommand\fastSSlonefinal{\texttt{SS-l1}}
\newcommand\fastSSTVfinal{\texttt{SS-TV}}

\newcommand\fastSSlonefinalProj{\texttt{SS-l1-decorr}}
\newcommand\fastSSTVfinalProj{\texttt{SS-TV-decorr}}

\newcommand\SSMethods{\textit{SS methods}}

\newtheorem{theorem}{Theorem}
\newtheorem{corollary}{Corollary}

\newtheorem{definition}{Definition}

\newtheorem{lemma}{Lemma}
\newtheorem{lemma*}{Lemma}
\newtheorem{remark}{Remark}

\newlength \myfigwidth
\newlength \myfigwidthll
\newlength \myfigwidthllI

\newlength \descriptionwidth

\if@twocolumn
\else
\fi
\if@twocolumn
\else
\fi

\if@twocolumn
\else
\fi

\begin{document}
%
\title{\LARGE{Compressive Source Separation:\\  Theory and Methods for Hyperspectral Imaging}}
%
%
\author{
Mohammad~Golbabaee$^*$~\IEEEmembership{Student Member~IEEE}\thanks{ $^*$ M. G. and S. A. equally contributed to this work.}, 
Simon~Arberet$^*$,
~and~Pierre~Vandergheynst
\thanks{The authors are with the Signal Processing Laboratory LTS2, Electrical Engineering Department, \'Ecole Polytechnique F\'ed\'erale de Lausanne (EPFL), Station 11, CH-1015 Lausanne, Switzerland.
This work was supported in part by the EU FET program through projects SMALL (FET-225913) and UNLocX (FET-255931), and the Swiss National Science Foundation under grant 200021-117884.
\protect\\
E-mail:\{mohammad.golbabaei,simon.arberet, pierre.vandergheynst\}@epfl.ch. }}
\maketitle

\begin{abstract}

With the development of numbers of high resolution data acquisition systems and the global requirement to lower the energy consumption, the development of efficient sensing techniques becomes critical. Recently, Compressed Sampling (CS) techniques, which exploit the sparsity of signals, have allowed to reconstruct signal and images with less measurements than the traditional Nyquist sensing approach. However, multichannel signals like Hyperspectral images (HSI) have additional structures, like inter-channel correlations, that are not taken into account in the classical CS scheme.

In this paper we exploit the linear mixture of sources model, that is the assumption that the multichannel signal is composed of a linear combination of sources, each of them having its own spectral signature, and propose new sampling schemes exploiting this model  to considerably decrease the number of measurements needed for the acquisition and source separation. Moreover, we give theoretical lower bounds on the number of measurements required to perform reconstruction of both the multichannel signal and its sources. We also proposed optimization algorithms and extensive experimentation on our target application which is HSI, and show that our approach recovers HSI with far less measurements and computational effort than traditional CS approaches.

\end{abstract}

\section{Introduction}

A Hyperspectral Image (HSI) is a collection of hundreds of images that have been acquired simultaneously in narrow and adjacent
spectral bands, typically by airborne sensors. 
HSI are produced by expensive spectrometers that sample the light reflected from a two-dimensional area. An HSI data set is thus a ``cube'' with two spatial and one spectral dimensions. Hyperspectral imagery has many applications including environmental monitoring, agriculture planning or mineral exploration. 
The  plurality of channels in HSI makes it possible to discriminate among  the various materials that make up a geographical area: each of them is represented by a unique spectral signature. Accordingly, HSI are often processed via clustering or source separation methods to obtain segmentation maps locating and labeling the various materials appearing in the image. Unfortunately, having multiple channels comes at a price: the sheer volume of data makes acquisition, transmission, storage and analysis of HSI computationally very challenging. Therefore, the problem addressed in this paper is to reduce the complexity of manipulating HSI via a suitable compression or dimensionality reduction technique. 


In this context  the emerging \textit{Compressive sensing} (CS) theory, which addresses the problem of recovering signals from few linear measurements, seems ideally suited~\cite{donoho2006compressed, CRT-stable2005}. The main assumption underlying CS is that the signal is sparse or compressible when expressed in a convenient basis. A signal $x \in \Rbb^n$ is said $k$-sparse in a basis $\Psi$ if it is a linear combination of only $k$ basis vectors of $\Psi$. 
The signal $x$ is said \textit{sparse} when $k \ll n$ and \textit{compressible} if the coefficient's magnitudes, when sorted, have a fast power-law decay, meaning that the signal has few large coefficients and many small coefficients. The recent literature abounds with examples of sparse models for signals and images.

While the CS-community has mostly focused on 1D or 2D signals, few works have been done on higher dimensional signals, in particular multi-array signals such as HSI. Extensions of wavelets basis for 3D data have been proposed \cite{duarte-kronecker} and rather generic sparse models have been exploited in \cite{SinglepixelHSI, CASSI} for designing innovative compressive hyperspectral imagers.
However, multi-array signals such as HSI have usually some structures that go beyond the sparsity assumption.
Indeed, HSI can be interpreted as a mixture of sources, each of them having a specific spectral signature. 
This model is widely used for unmixing HSI \cite{keshava2002spectral, nascimento2005vertex,1677768,1261124,arberet2010hyper}, that is extracting, form the HSI, each  source and their respective spectral signatures. 

The main focus of this paper is to exploit, beyond the sparsity assumption,  
an additional structured model, the \emph{linear mixture} model, so as to reconstruct
and separate the sources of multi-array signals assuming we know their spectra (or mixing parameters) as side information. Note that this hypothesis is validated in many applications where the elements or materials composing the data are known and their spectra tabulated.
This idea was first introduced in two of our conference papers \cite{golbabaee2010CSBSS, golbabaee2010distributed}.
In this paper, we introduce and analyze a new sampling scheme, which exploits this structured model, and 
that has the following important properties:
\begin{itemize}   
\item the number of measurements, or samples, does not scale with the number of channels,
\item the recovery results do not depend on the conditioning of the mixing matrix (as long as the mixing spectra are linearly independent).
\end{itemize}
We propose new algorithms for HSI \emph{compressive source separation} (CSS), that is source separation and data reconstruction from compressed measurements, which are based on exploiting the linear mixture structure and TV, $\ell_1$ or $\ell_0$ regularization. We establish that sources can be efficiently separated directly on the compressed measurements, i.e avoiding to run a source separation algorithm on this high-dimensional raw data, thereby eliminating this important bottleneck and providing a rather striking example of compressed domain data processing.
We provide theoretical guaranties  and intensive experiments which show that, with this approach, we can reconstruct a multi-array signal from compressed measurements with a far better accuracy than traditional CS approaches. For example, we are able to reconstruct HSI datasets with only $3\%$ relative error from $3\%$ of measurements and less than $1\permil$ of data transmission, with an algorithm that is more than $40$ times faster. 
While the main target application of this paper is HSI, our model and the theoretical analysis is general and could be applied to other multi-array signals like e.g. Positive Emission Tomography (PET) or distributed sensing. 


The remainder of this paper is structured as follows. The necessary background and notations are first introduced in Section~\ref{sec-bgd}. We then propose, in Section \ref{ch5-sec:sampling}, two acquisition schemes that exploit the prior knowledge of the mixing parameters so as to perform a decorrelation step. In Section \ref{ch5-sec: th bounds}, we provide theoretical guarantees for both source identification and data reconstruction.  We determine the number of CS measurements sufficient for robust source identification and signal reconstruction as a function of the sparsity of the sources, sampling SNR and the conditioning of their corresponding mixture parameters. 
In Section \ref{ch5-app} we discuss in further details the application of our acquisition and recovery schemes for HSI.  We introduce different recovery algorithms that we compare with the classical methods, for various CS acquisition schemes on two sets of HSI. Finally, in the spirit of reproducible research, the code and data needed to reproduce the experimental sections of this paper is openly available at \url{http://infoscience.epfl.ch/record/180911}.

\section{Background and Notations}
\label{sec-bgd}
\subsection{CS of Multichannel Signals\label{sec_CS_basics}}

We represent a multichannel signal with a matrix $X \in \Rbb^{\na \times \nb}$  where $\nb$ is the number of channels and $\na$ is the dimension of signal in each channel.
The CS acquisition protocol of a multichannel signal $X$ is a linear mapping $\acal: \Rbb^{\na \times  \nb}\rightarrow \Rbb^m$ of $X$ into a CS measurement vector $y\in \Rbb^{m}$ contaminated by the measurement noise $z\in \Rbb^{m}$: 
$$y = \acal(X)+z.$$
When $m\ll \na\nb$ the signal is effectively compressed.
The main goal of CS is to recover the signal $X$ from the fewest amount of measurements $m$.
Note that any linear mapping $\acal(X)$ can be written in matrix form $A X_{vec} := \acal(X)$, where $A \in \Rbb^{m \times \na\nb}$ and  $X_{vec} \in  \Rbb^{\na\nb}$ is the vectorized form of matrix $X \in \Rbb^{\na \times \nb}$:
\begin{equation}\label{sampling model2}
y = AX_{vec}+z.
\end{equation}

In order to recover $X_{vec}$,  we can search for the sparsest vector $X_{vec}$ which is consistent with the measurement error, 
leading to the following $\ell_0$-minimization problem:
\begin{eqnarray}\label{eq_L0_NA}
\argmin_{X_{vec}} \|X_{vec}\|_{\ell_0} \qquad s.t. \qquad \|y - AX_{vec} \|_{2} \leq \varepsilon,
\end{eqnarray}
where $\varepsilon $ is an upper bound on the norm of the noise vector (i.e. $ \|z\|_{2}\leq\varepsilon$), $\| \cdot \|_{\ell_{0}}$ denotes the $\ell_{0}$ quasi-norm of a vector (\ie, the number of its nonzero coefficients).
Unfortunately, this combinatorial minimization problem is NP-hard in general \cite{mallat1993matching,Natarajan_1995}.
However, there are two tractable alternatives to solve problem \eqref{eq_L0_NA}: 
The convex relaxation leading to $\ell_1$-minimization and greedy algorithms such as matching pursuits (MP) \cite{mallat1993matching} or Iterative Hard Thresholding (IHT) \cite{IHT-sparse}.
Both types of approaches provide conditions on the matrix $A$ and on the sparsity $k$ such that the recovered solution coincides
with the original signal $X_{vec}$, and consequently also with the solution of  \eqref{eq_L0_NA}.

The $\ell_1$ minimization approach consists in solving the following non-smooth convex optimization problem called Basis Pursuit DeNoising (BPDN):
\begin{eqnarray}\label{eq_BPDN_NA}
\argmin_{X_{vec}} \|X_{vec}\|_{1} \qquad s.t. \qquad \|y - AX_{vec} \|_{2} \leq \varepsilon,
\end{eqnarray}
where $\|\cdot\|_{1}$ denotes the $\ell_1$ norm, which is equal to the sum of the
absolute values of the vector entries, $\|\cdot\|_{2}$ denotes the $\ell_2$ or Euclidean norm.

It has been shown in \cite{donoho2006compressed, CRT-stable2005, candes2008restricted} that approximating the sparse recovery problem  by the $\ell_{1}$ minimization \eqref{eq_BPDN_NA} can stably recover the $kn_2$-sparse original solution (i.e. $k$-sparse signal per channel) whenever $A$ satisfies the so-called \emph{restricted isometry property} (RIP). 
This result guarantees that sparse signals can be perfectly recovered from noise-free measurements and that the recovery
process is robust to the presence of noise.
The computation of the isometry constants for a given matrix is prohibitive in practice, but certain classes of matrices, such as matrices with 
independent Gaussian or Bernoulli entries, obey the RIP condition with high probability (see Theorem 5.2 in \cite{JL-lemma}) as long as:
\begin{eqnarray}\label{ch5-meas bound}
m \geq c\, n_2k\log(\na/k).
\end{eqnarray}
for a fixed constant $c$.



\subsection{Sparse Regularization of a Multichannel Signal\label{sec_sparse_reg_MultArra_sig}}

Usually the data $X_{vec}$ is not directly sparse, but sparse in a basis $\Psin \in \Rbb^{\na\nb \times \na\nb}$.
In that case, the $\ell_1$ regularization approach consists in solving the following problem which generalizes problem 
\eqref{eq_BPDN_NA}:
\begin{eqnarray}\label{eq_BPDN_NA_Basis}
\argmin_{\Thetan_{vec}} \| \Thetan_{vec}\|_{1} \qquad s.t. \qquad \|y - A \Psin \Thetan_{vec} \|_{2} \leq \varepsilon,
\end{eqnarray}
with $X_{vec} = \Psin \Thetan_{vec}$.
Stable reconstruction by solving problem \eqref{eq_BPDN_NA_Basis} 
 is guaranteed as long as the $A \Psin$ matrix satisfies the RIP.
 When the data is a multichannel image, a classical basis is a block diagonal orthonormal basis
  $\Psin = \Id_{\nb}\otimes \Psia$ \footnote{$\Id_{\nb}$ is the $\nb\times\nb$ identity matrix and $\otimes$ denotes the matrix Kronecker product.} where $\Psia \in \Rbb^{\na\times \na}$ denotes a proper 2-dimensional wavelet basis.

Another classical approach to regularize the data (specially images) is the 
total variation (TV) penalty \cite{TV-ROF}, which tends to generate images with piecewise smooth regions and sharp boundaries.
Replacing the $\ell_1$ norm with the $TV$ norm on each channel $X_{j}$ of the multichannel in problem \eqref{eq_BPDN_NA_Basis} leads to the Total Variation De-Noising (TVDN) problem:
\begin{eqnarray}\label{eq_BPDN_NA_TV}
\argmin_{X}  \sum_{j=1}^{\nb}\|X_{j}\|_{TV}  \qquad s.t. \qquad \|y - A X_{vec} \|_{2} \leq \varepsilon.
\end{eqnarray}

\subsection{The Linear Mixture Model}

One of the most practical setups of a multichannel signal is when the multichannel data matrix  $X$ is derived by a \emph{sparse linear mixture} model as follows: 
\begin{eqnarray}\label{mix model}
X = \sbf \mixMatrix^{T}.
\end{eqnarray}
Here, $\sbf \in \Rbb^{\na \times \rho}$ denotes the \emph{source matrix} whose $i$th column contains the proportion of the source $i$ at each pixel.  Each source is mixed with the corresponding column of the \emph{mixing matrix} $\mixMatrix \in \Rbb^{\nb \times \rho}$ in order to generate the full multichannel data. Each column of $\mixMatrix$ contains the spectrum of the corresponding source. The observed signal in any channel $j \in  \{1,\ldots, \nb\}$ is thus a linear combination of $\rho$ source signals:
\[
X_{j} = \sum_{i = 1}^{\rho} [\mixMatrix]_{j,i}\,\sbf_{i}. 
\]

\subsection{Mixing Parameters as Side Information for Multichannel CS Recovery \label{sec_mix_par_side_Inf_MACSR}}
In certain multichannel signal acquisition setups the mixing parameters $\mixMatrix$ are known at both decoder and encoder sides. In particular, this is the case in many remote sensing applications where the spectra of common materials are tabulated. Such prior efficiently restricts the degrees of freedom of the entire data matrix to the sparse coefficients of the underlying sources. 
Indeed, we will show that, when we know the mixing parameters $\mixMatrix$, the inverse problem consisting in recovering the multichannel signal $X$ from the measurements $y$ in \eqref{sampling model2} is equivalent to the problem of recovering the sources $\sbf_{vec}$ from the following measurements:
\begin{equation}\label{sampling model SS}
y = A\Phi  \sbf_{vec}+z,
\end{equation}
\noindent with $\Phi = \mixMatrix \otimes \Id_{\na}$.
The source coefficients can then be recovered  by solving a convex optimization problem such as \eqref{eq_BPDN_NA_Basis}, where $A$ is replaced by $A\Phi$ and the multichannel signal can be reconstructed by applying the mixing matrix to the recovered source matrix according to the linear mixture model \eqref{mix model}.
This approach has the advantage of solving two problems: i) source separation directly from the compressive measurements
, ii) data compressive sampling via source separation or, equivalently, via a particular structured sparse model.

%

\section{Compressive Multichannel Signal Acquisition Schemes}\label{ch5-sec:sampling}

If the multichannel signal follows the linear mixture model \eqref{mix model}, the knowledge of the mixing matrix can be used efficiently.
The sparse source coefficients  can be directly recovered from the measurements. 
In this section we introduce a decorrelation mechanism, applied at the acquisition process or as a post-processing step,
which has two main advantages:
first it leads to strong dimensionality reduction and secondly it improves the conditioning of the recovery problem.

\subsection{Multichannel Recovery via Source Recovery\label{sec_l1_recovery_scheme}}

When we know the mixing matrix $\mixMatrix$, and  thanks to the property  ($(BCD)_{vec}=(D^{T}\otimes B)C_{vec}$) of the Kronecker product, the sampling equation \eqref{sampling model2} (in the noise free case) can be written as: 
\begin{equation}
A X_{vec} = A(\sbf\mixMatrix^{T})_{vec} = A  \underbrace{ (\mixMatrix \otimes \Id_{\na})}_{ \defeq \Phi}  \sbf_{vec} = A  \Phi  \sbf_{vec}.
\end{equation}

Then, the $\ell_1$ regularization approach for the recovery of the whole data consists in finding the sparsest coefficients vector $\Thetabf_{vec} \in \Rbb^{\rho \na}$ of the sources vector $\sbf_{vec}=\Psibf \Thetabf_{vec}$ in a basis $\Psibf \in \Rbb^{\rho\na \times \rho\na}$, where e.g. 
 $\Psibf = \Id_{\rho}\otimes \Psia$ is a block diagonal orthonormal basis, 
   through the following minimization:
\begin{eqnarray}\label{l1 SSCS}
\argmin_{\Thetabf_{vec}} \|\Thetabf_{vec}\|_{1} \qquad s.t. \qquad \|y - A \Phi \Psibf \Thetabf_{vec} \|_{2} \leq \varepsilon.
\end{eqnarray}

This corresponds to a ``synthesis'' formulation of BPDN using a basis $\Psibf$.
The ``analysis''  formulation, which is equivalent to the synthesis one when $\Psibf$ is a basis but different when $\Psibf$ is a redundant dictionary, 
 consists in solving the following problem with respect to the sources instead of its coefficients:
 \begin{eqnarray}\label{l1 SSCS analysis}
\argmin_{\sbf_{vec}} \| \Psibf^* \sbf_{vec}\|_{1} \qquad s.t. \qquad \|y - A \Phi \sbf_{vec} \|_{2} \leq \varepsilon,
\end{eqnarray}
where $\Psibf^*$ is the adjoint of the operator $\Psibf $.

The data $X$ can then be recovered via the mixture model $\widehat  X = \widehat  \sbf \mixMatrix^{T}$, with $\widehat  \sbf_{vec}$ being either the solution of the analysis problem   \eqref{l1 SSCS analysis} or $\widehat  \sbf_{vec}$ being equal to 
$\widehat \sbf_{vec}=\Psibf  \widehat \Thetabf_{vec}$ with $\widehat \Thetabf_{vec}$, solution of the synthesis problem \eqref{l1 SSCS}.


\subsection{Decorrelation Scheme \label{sec_decorr_sect}}

We have seen in section \ref{sec_CS_basics}, that the conditions to recover the signal from the noisy measurements $y = AX_{vec}+z$ depend on properties (such as RIP) of the sensing matrix $A$.
We introduce a particular structure for the sampling matrix $A$ which benefits from the available knowledge of the mixture parameters $\mixMatrix$ and incorporates data decorrelation into the compressive acquisition.

\subsubsection{Decorrelating Multichannel CS Acquisition}\label{ch5-sec: dense decorr sampling}

The decorrelation mechanism consists of applying the Moore-Penrose pseudo inverse matrix $\mixMatrix^{\dag} = (\mixMatrix^{T}\mixMatrix)^{-1} \mixMatrix^T$ in order  to remove the underlying dependencies among CS measurements. We therefore propose the following sampling matrix: 
\begin{eqnarray}
A = \mixMatrix^{\dag}\otimes \atilde.\label{decorr matrix}
\end{eqnarray}
The main sampling matrix is generated from a smaller-size $\mhat \times \na$ \emph{core sampling matrix} $\atilde$. Note that CS imposes $\mhat \ll \na$. 

%

The total number of measurements is $m = \rho\,\mhat$. Applying the sampling matrix $A$ of \eqref{decorr matrix} on multichannel data results in the following CS measurements: 
 \begin{eqnarray}
 y 
 &=& A \Phi  \sbf_{vec} +z\\
&=&  \underbrace{(\mixMatrix^{\dag}\otimes \atilde)}_{A}  \underbrace{(\mixMatrix \otimes \Id_{\na})}_{\Phi} \, \sbf_{vec} +z, \nonumber \\
&=& \underbrace{ (\Id_{\rho} \otimes \atilde)}_{\defeq \usm[\rho]} \, \sbf_{vec}+z.  \label{eq_simpl_decorrA}
\label{decorr sampling}
 \end{eqnarray}

 
 
 The third equality comes from the following property: $(B \otimes C)(D \otimes F) = BD\otimes CF$, and 
 $\usm[\rho]$ is a block diagonal matrix whose $\rho$ diagonal blocks are populated with $\atilde$:
 $
 \usm[\rho] \defeq \Id_{\rho}\otimes \,\atilde.
 $
  

 As we can observe in \eqref{decorr sampling} and thanks to the specific structure of the sampling matrix, the mixing parameters $\mixMatrix$ are discarded from the formulation and each source (each column of $\sbf$) is directly subsampled by the same matrix $\atilde$.
 

 \subsubsection{Uniform Multichannel CS Acquisition}\label{ch5-sec: Uniform sampling}
 
 In many practical setups the acquisition scheme can not be arbitrarily chosen and is rather determined by various constraints posed by the physics of the signals and the implementation technology. Certain acquisition systems such as Rice's single-pixel hyperspectral imager \cite{SinglepixelHSI} are using a universal random matrix to sample independently data in each channel. In this case, 
 acquisition models such as \eqref{decorr matrix}, which require inter-channel interactions for compressed sampling, simply cannot be implemented. Here, the sampling matrix $A$ in \eqref{sampling model2} is block diagonal with $\nb$ blocks (each applies on a certain channel) that are populated by a unique $\mhat \times \na$ matrix (similarly as $\atilde$ in \eqref{eq_simpl_decorrA}): 
 \begin{equation}\label{eq:uniform sensing matrix}
A = \usm[\nb] \defeq  \Id_{\nb}\otimes \,\atilde.
\end{equation}
The total number of measurements is then $m=\nb\,\mhat$. 
Reshaping $y$ and $z$ correspondingly into $\mhat \times \nb$ matrices $Y$ (the measurement matrix) and $Z$ (the noise matrix) leads to the following equivalent formulation:
\[
Y = \atilde X +Z.
\]
\subsubsection{Decorrelation-based Uniform Sampling}\label{ch5-sec: uniform decorr sampling}

 A decorrelation step similar to the one introduced in Section \ref{ch5-sec: dense decorr sampling} can be applied on the CS measurements. It consists in multiplying the rows of the measurement matrix by $(\mixMatrix^{\dag})^T$ and reducing the dimensionality of $ Y$ to an $\mhat \times \rho$ matrix as follows:  
 \begin{eqnarray}
 Y^{*} &=& Y (\mixMatrix^{\dag})^T \nonumber \\
 &=& \atilde\, \sbf + Z^{*}, \nonumber
 \end{eqnarray} 
where, $Z^{*} = Z (\mixMatrix^{\dag})^T$. 
By reshaping $Y^{*}$ and $Z^{*}$ into the vectors $y^{*}$ and $z^{*}$, we can observe that the outcome of such \emph{decorrelation-based uniform sampling} leads to an expression similar to  \eqref{decorr sampling} \ie,
\begin{eqnarray}\label{proj sampling}
y^{*}= \usm[\rho]\sbf_{vec} + z^{*}.
\end{eqnarray}

This decorrelating scheme favorably  reduces the dimension of the data: at the acquisition stage, the  total number of samples is $\nb\, \mhat$ but at the
transmission and decoding stages the number of samples is only $\rho\, \mhat \ll \nb\, \mhat$.



\label{ch5-sec: recovery}

For the \emph{decorrelating} sampling schemes described in section \ref{ch5-sec: dense decorr sampling} and \ref{ch5-sec: uniform decorr sampling},  
the $\ell_{1}$ minimization (e.g. the "synthesis" problem \eqref{l1 SSCS}) of section \ref{sec_l1_recovery_scheme}  takes the following form:
\begin{eqnarray}\label{l1 SSCS1}
\argmin_{\Thetabf_{vec}} \|\Thetabf_{vec}\|_{1} \qquad s.t. \qquad \|y - \usm[\rho] \,\Psibf \Thetabf_{vec} \|_{2} \leq \varepsilon,
\end{eqnarray}
which, in the noiseless case can be decoupled into $\rho$ independent $\ell_{1}$ minimizations, each of them corresponding to a certain source compressed by a universal matrix $\atilde$. 
In Section \ref{ch5-sec: th bounds} we provide the theoretical analysis of such recovery scheme for various acquisition schemes. 

\section{Main Theoretical Analysis}\label{ch5-sec: th bounds}

Compressive sparse source recovery is closely related to the problem of compressed sensing with \emph{redundant dictionaries} \cite{DicCS, Candes-Eldar}. Indeed, the later problem has the same formulation as in \eqref{l1 SSCS} by replacing $\Phi$ by an overcomplete dictionary matrix.
The first part of this section provides an overview of the CS literature on redundant dictionaries. In the second part, we derive new performance bounds that extend the former results for a larger class of dictionaries. In the third part, we cast the sparse source separation problem as a particular case of CS recovery using redundant dictionaries and we give a bound on the performance of the $\ell_{1}$ minimization for each of  the considered CS acquisition schemes (dense, uniform and decorrelated). 

\subsection{Compressed Sensing and Redundant Dictionaries}\label{ch5-sec: RIP dic}

 Let $x\in \Rbb^{n}$ be a vector that is sparse in a  dictionary $\D \in \Rbb ^{n\times d}$ (\ie, $x = \D\,\theta$ with, $\theta \in \Rbb^{d}$).
The $\ell_{1}$  minimization approach for recovering $\theta$ (equivalently $x$) from the compressive measurements $y=Ax+z$ consists in solving: 
 \begin{eqnarray}\label{dic CS}
\argmin_{\theta} \|\theta\|_{1} \qquad s.t. \qquad \|y - A\D \theta \|_{2} \leq \varepsilon,
\end{eqnarray}
where, $\|z\|_{2}\leq \varepsilon$. Note that in this section $A$ is a sampling matrix of size $m\times n$ and 
the dictionary $\D$ typically contains a large number of columns ($d\gg n$).

It has been shown in \cite{donoho2006compressed, CRT-stable2005} that the $\ell_{1}$ minimization \eqref{dic CS} can stably recover the original solution whenever $A \D$ satisfies the \emph{restricted isometry property} (RIP). More precisely, if for all $k$-sparse vectors $\theta$ the following RIP property holds:
\begin{equation}\label{ch5-RIP classic}
 \Big(1-\delta_{k}(A\D) \Big) \|\theta\|^{2}_{2} \leq \|A\D \,\theta\|^{2}_{2} \leq \Big(1+\delta_{k}(A\D) \Big) \|\theta\|^{2}_{2}
\end{equation}
with the RIP constant of order $k$, $\delta_{k}(A\D) \leq \sqrt{2}-1$, then the solution $\widehat \theta$ to \eqref{dic CS} satisfies the following error bound:
\begin{eqnarray}\label{dic error bound}
\| \theta - \widehat \theta \|_{2} \leq c_{0} \,k^{-1/2}\, \|\theta - \theta_{k}\|_{1}+ c_{1} \varepsilon,
\end{eqnarray}
for some positive constants $c_{0}, c_{1}$, and where $\theta_{k}$ is the best $k$-sparse approximation of $\theta$. Now the question is how many CS measurements are sufficient so that $A\D$ satisfies the RIP ?
It has been shown in \cite{DicCS} that, for a certain class of random sampling matrices $A$ (\eg, with i.i.d. Gaussian, Bernoulli or subgaussian elements), with very high probability the RIP constant $\delta_{k}(A\D)$ is bounded by:
\begin{eqnarray}\label{RIP dic}
\delta_{k}(A\D) \leq \delta_{k}(A)+\delta_{k}(\D)+\delta_{k}(A)\delta_{k}(\D).
\end{eqnarray}

If $\D$ is an orthonormal basis, then $\delta_{k}(\D)=0$ and $A\D$ becomes another subgaussian matrix with a similar distribution as for $A$ and thus \eqref{RIP dic} holds with equality \ie, $\delta_{k}(A\D)=\delta_{k}(A)$. 

Considering the recovery condition using $\ell_{1}$ minimization (\ie, $\delta_{k}(A\D)\leq \sqrt2-1$) and the bound in \eqref{RIP dic}, we can conclude that $A$ must satisfy RIP with the following constant:
\begin{eqnarray} \label{RIP dic2}
 \delta_{k}(A) \leq \frac{ \sqrt 2-1-\delta_{k}(\D)} {1+\delta_{k}(\D)}.
\end{eqnarray}

Moreover, using the Johnson-Lindenstrauss lemma, it has been shown that (see Theorem 5.2 in \cite{JL-lemma}) a random matrix $A$ whose elements are drawn independently at random from  Gaussian, Bernoulli or subgaussian distributions satisfies RIP as long as we have:
\begin{eqnarray}\label{ch5-meas bound}
m \geq c\, k\log(n/k),
\end{eqnarray}
for a constant $c$ depending on the RIP constant of $A$ \ie, the higher $\delta_{k}(A)$, the smaller $c$. If $\D$ is not a unitary matrix, $\delta_{k}(\D)$ becomes a positive constant and the more coherent the columns of $\D$, the larger its RIP constant. 
Therefore, there is a tradeoff for compressed sensing using redundant dictionaries: redundancy can result in a more compact representations of the signals \ie, smaller $k$, and thus less measurements are required for CS recovery using \eqref{dic CS}. Meanwhile, too much redundancy can lead to an awfully large constant in \eqref{ch5-meas bound} implying that more CS measurements are required to overcome the uncertainties brought by over completeness.

\subsection{Performance Bounds for Compressed Sensing using Asymmetric-RIP Dictionaries}\label{ch5-ripless guaranty}

In Section \ref{ch5-SStheory} we will show that applying the classical RIP based analysis 
results in conditions that are too restrictive to guaranty the source recovery.
Therefore in this part and in order to overcome such limitations, we derive a new theoretical performance bound that uses different notions of RIP.
We begin by introducing the notions of the \emph{asymmetric restricted isometry property} (A-RIP) and the \emph{restricted condition number} of a dictionary $\D$.

\begin{definition}\label{ch5-def1}
For a positive integer $k\in \Nbb$ , an  $n\times d$ matrix $\D$ satisfies the asymmetric restricted isometry property, if for all $k$-sparse $x\in \Rbb^{d}$ the following inequalities hold:
\begin{eqnarray}
\lo_{k}(\D) \| x\|_{2} \leq \|\D x\|_{2} \leq \up_{k}(\D)\| x\|_{2},
\end{eqnarray}
\noindent where, $\lo_{k}(\D)$ and $\up_{k}(\D)$ are correspondingly the largest and the smallest constants for which the inequalities above hold. The restricted condition number of $\D$ is defined as:
\begin{eqnarray}
\xi_{k}(\D) \defeq \frac{ \up_{k}(\D) } {\lo_{k}(\D)}.
\end{eqnarray}
\end{definition}

In addition, we use a different notion of RIP for the compression matrix $A$, namely, the \emph{Dictionary Restricted Isometry Property} (D-RIP), proposed by Candes \etal in \cite{Candes-Eldar}: 

\begin{definition}\label{ch5-def2}
For a positive integer $k \in \Nbb$, a matrix $A$ satisfies the D-RIP  adapted to a dictionary $\D$ as long as for all $k$-sparse vectors $x$ the following inequalities hold:
\begin{eqnarray}
(1-\delta^{*}_{k})\| \D x\|_{2}^{2} \leq \|A\D x\|_{2}^{2} \leq (1+\delta^{*}_{k})\| \D x\|_{2}^{2}.
\end{eqnarray}
The D-RIP constant $\delta^{*}_{k}$ is the smallest constant for which the property above holds.
\end{definition}

This definition extends the classical RIP (which deals with signals that are sparse in the canonical basis) to linear mappings that are able to stably embed all low dimensional subspaces spanned by every $k$ columns of a redundant dictionary $\D$. 

As in \cite{Candes-Eldar}, we suppose that $A$ is an $m \times n$ matrix drawn at random from certain distributions that satisfy the following concentration bound for any vector $x$: 
\begin{eqnarray}\label{ch5: concentration}
\mathbf{Pr}  \left(  \,  \left|  \| A x \|_{2}^{2} - \|x\|^{2}_{2} \right| > t \|x\|_{2}^{2} \,  \right) \leq C \exp \left( -c \, m\right),
\end{eqnarray}
for some constants $C$ and $c >0$ that are only depending on $t$. Then, $A$ will satisfy the D-RIP for \emph{any} $n\times d$ dictionary $\D$ with overwhelming probability if
\begin{eqnarray}
m \gtrsim \O(k\log(d/k)). \nonumber
\end{eqnarray}


\begin{remark}\label{ch5-remark1}
Matrices $A\in \Rbb^{m\times n}$ whose elements are independently drawn at random from  Gaussian, Bernoulli or (in general) subgaussian distributions satisfy the concentration bound in \eqref{ch5: concentration} and therefore satisfy D-RIP for any $n\times d$ dictionary as long as $m \gtrsim \O(k\log(d/k))$.
\end{remark}

Based on these definitions we establish the following theorem in order to bound the performance of the $\ell_{1}$ minimization in \eqref{dic CS}:

\begin{theorem}\label{thSS}
Given a matrix $A$ that satisfies the D-RIP adapted to a dictionary $\D$, with the constant $\delta^{*}_{\gamma k}< 1/3$ where $\gamma \geq 1+2\xi^{2}_{\gamma k}(\D)$, then the solution $\widehat \theta$ to \eqref{dic CS} obeys the following bound:
\begin{eqnarray}\label{error bound SS}
\| \theta - \widehat \theta \|_{2} \leq c'_{0} \, k^{-1/2} \, \|\theta - \theta_{k}\|_{1} + c'_{1} \varepsilon,
\end{eqnarray}
for some positive constants $c'_{0}, c'_{1}$.
\end{theorem}

The proof of this theorem is given in Appendix.
 Using Remark \ref{ch5-remark1}, the following result is straightforward: 

\begin{corollary}\label{ch5-corollay1}
For $A$ whose elements are drawn independently at random from Gaussian, Bernoulli or subgaussian distributions, the solution to \eqref{dic CS} obeys the error bound \eqref{error bound SS} with an overwhelming probability and for any dictionary with a finite Restricted Condition Number $\xi_{\gamma k}(\D)$, if 
\begin{eqnarray}\label{DRIP meas}
m \gtrsim \gamma k \,\log{(d /\gamma k)} .
\end{eqnarray}
\end{corollary}

Comparing to the bound \eqref{ch5-meas bound} based on the classical RIP analysis, we see that \eqref{DRIP meas} features the same scaling-order for the number of measurements. In addition, for both types of analysis the constant factors grow as the atoms of the dictionary become more coherent and therefore, more CS measurement are required. 

Note that this result requires neither $A\D$ nor the dictionary $\D$ to satisfy the \emph{classical} RIP. 
In the next section, we apply these results to guaranty the performance of the $\ell_{1}$ minimization approach \eqref{l1 SSCS} for source identification and in particular, for the case  where $\hbf$ is not well-conditioned.

\subsection{Theoretical Guaranties for Source Recovery using $\ell_{1}$ Minimization}\label{ch5-SStheory}

Sparse source recovery from  compressive measurements using  $\ell_{1}$ minimization  \eqref{l1 SSCS} is a particular case of the compressed sensing problem using dictionaries \eqref{dic CS}. Indeed, for the source recovery problem, $\theta$ and the dictionary matrix $\D$ are  replaced respectively with $\Thetabf_{vec}$ and $\Phi' \defeq \Phi \Psibf =(\mixMatrix \otimes \Id_{\na}) \Psibf $,  and consequently, $n = \na\nb$ and $d=\rho\na$. The only difference here is that $\Phi' $ is a tall matrix (\ie, $d \leq n$) due to its specific construction and the assumption of having few number of sources (\ie, $\rho \leq \nb$). Though there is no redundancy in $\Phi' $ in terms of the number of columns, there is uncertainty at the sparse decoder because of \emph{coherent} columns. The following lemma which has been proven in \cite{duarte-kronecker} (see Lemma 2 in \cite{duarte-kronecker}) shows that the conditioning of $\Phi'  $ is directly related to the conditioning of the underlying mixture parameters \ie, intuitively, if the columns of $\mixMatrix$ become coherent, so become the columns of $\Phi'$. 

\begin{lemma}\label{RIP kron}
For matrices $V_{1}, V_{2}, \ldots, V_{\ell}$ with restricted isometry constants $\delta_{k}(V_{1}), \delta_{k}(V_{1}),\ldots, \delta_{k}(V_{\ell})$ respectively, we have:
\begin{eqnarray}\label{kron RIP}
\delta_{k}(V_{1}\otimes V_{2}\otimes \ldots \otimes V_{\ell}) \leq \prod_{i=1}^{\ell} \Big(1+\delta_{k}(V_{i})\Big) -1.
\end{eqnarray}
\end{lemma}


Since the RIP constant of any orthonormal basis is zero (\eg, $\delta_{k}(\Id_{\na})=0$), and since 
$\Psibf$ is an orthogonal matrix, we can deduce the following bound on the RIP constant 
  of $\Phi' = (\mixMatrix \otimes \Id_{\na}) \Psibf$ by applying Lemma \ref{RIP kron}:
\begin{eqnarray}
\delta_{k}(\Phi')&=& \delta_{k}(\Phi) \nonumber\\
&\leq& \delta_{k}(\mixMatrix)  \label{ch5-eq1} \\
&\leq& \eta \defeq \max \Big( 1-\sigma^{2}_{\min}(\mixMatrix),\,\, \sigma^{2}_{\max}(\mixMatrix)-1 \Big).  \label{ch5-eq2}
\end{eqnarray}
For $k\leq \rho$ one can use \eqref{ch5-eq1} (which then holds with equality), and more generally \eqref{ch5-eq2} for any $k$. Note that \eqref{ch5-eq2} follows by the definition of the RIP constant and it only holds if $\mixMatrix$ is properly normalized so that $1\leq\sigma_{\max}(\mixMatrix)<2$ and $ 0< \sigma_{\min}(\mixMatrix)\leq 1$. \footnote{This can be done by dividing $\mixMatrix$ and multiplying $\sbf$  by $\big(\sigma_{\max}(\mixMatrix)+\sigma_{\min}(\mixMatrix) \big)/2$, respectively. } 

Moreover, due to the properties of the extreme singular values of the Kronecker product of two matrices:
\[
\sigma_{\max}(V_{1}\otimes V_{2}) = \sigma_{\max}(V_{1})\, \sigma_{\max}(V_{2}),\]
\[
\sigma_{\min}(V_{1}\otimes V_{2}) = \sigma_{\min}(V_{1})\, \sigma_{\min}(V_{2}),
\]
and according to  Definition \ref{ch5-def1}, we can bound the restricted condition number of $\Phi'$ as follows:
\begin{eqnarray}\label{RCN SS}
\xi_{k}(\Phi') \leq \frac{ \sigma_{\max}(\Phi') } {\sigma_{\min}(\Phi') } = \frac{ \sigma_{\max}(\mixMatrix) } {\sigma_{\min}(\mixMatrix) } \defeq \xi(\mixMatrix),
\end{eqnarray}
where, $\xi(.)$ (without subscript) denotes the standard definition of the condition number of a matrix.
With those descriptions, the performance of the sparse source recovery using \eqref{l1 SSCS} can be easily characterized by any of the  previous types of performance bound of sections \ref{ch5-sec: RIP dic} and \ref{ch5-ripless guaranty}. 

According to the standard definition of the RIP for the matrix $\Phi'$, we can bound its restricted condition number $\xi_{k}(\Phi')$ as follows:
\[
\xi_{k}(\Phi') \leq \sqrt \frac{1+\delta_{k}(\Phi')}{1-\delta_{k}(\Phi') }.
\]

Recall that, the classical RIP based analysis in section \ref{ch5-sec: RIP dic} requires $\delta_{k}(\Phi')<\sqrt 2 -1$ (in order to have $\delta_{k}(A)>0$ in \eqref{RIP dic2}), which implies $\xi_{k}(\Phi') < \sqrt{\sqrt 2+1}$, or consequently $\xi(\mixMatrix) < \sqrt{\sqrt 2+1}$. 
This severely restricts the application of such analysis for a limited class of relatively well-conditioned mixture parameters. 

To address this limitation, we use the second theoretical analysis based on the D-RIP of the compression matrix presented in section \ref{ch5-ripless guaranty}. The following theorem is a corollary of Theorem \ref{thSS}:

\begin{theorem}\label{ch5-th2}
Given a mixture matrix $\mixMatrix$ whose condition number is $\xi(\mixMatrix)$, and a matrix $A$ that satisfies the D-RIP adapted to $\mixMatrix \otimes \Id_{\na}$ with the constant $\delta^{*}_{\gamma' k}< 1/3$ where $\gamma' = 1+2\xi^{2}(\mixMatrix)$, then the solution $\widehat \Thetabf_{vec}$ to \eqref{l1 SSCS} obeys the following bound for the same constants $c'_{0}, c'_{1}$ as in \eqref{error bound SS}:
\begin{eqnarray}\label{error bound SS2}
 \| \Thetabf_{vec} - \widehat \Thetabf_{vec} \|_{2} \leq c'_{0} \,k^{-1/2}  \|\Thetabf_{vec} - (\Thetabf_{vec})_{k}\|_{1} + c'_{1} \varepsilon.
 \end{eqnarray}
\end{theorem}

Comparing to Theorem \ref{thSS}, $\D$ is replaced by $\Phi'$ and $\gamma$ is set to $\gamma'$  which satisfies the requirement of Theorem \ref{thSS} \ie, according to \eqref{RCN SS} we have $\gamma' \geq 1+2\xi_{\gamma' k}^{2}(\mixMatrix)$. As we can see, this analysis is valid for a much wider range of condition number namely, $\xi(\mixMatrix) \leq \sqrt \frac{\na\nb/k-1}{2}$. \footnote{As for $\gamma'k\geq \na\nb$ an $\na\nb\times \na\nb$ identity matrix $A$ always satisfies $\delta^*_{\gamma'k}=0$ (\ie there is no advantage by replacing the full Nyquist sampling with CS), Theorem \ref{ch5-th2} becomes useful only when we have $\gamma'k < \na\nb$ which for the value of $\gamma'$ in the theorem implies $\xi(\mixMatrix) \leq \sqrt \frac{\na\nb/k-1}{2}$.}


 Now, if we use this approximation to recover the multichannel data \ie, $\widehat X = \widehat \sbf\mixMatrix^{T}$, 
  the reconstruction error can be bounded using \eqref{error bound SS2} and the following inequality:
\begin{eqnarray}\label{rec-err bound}
\|X-\widehat X\|_{F} &\leq& \sigma_{\max}(\mixMatrix)\|\sbf-\widehat \sbf \|_{F} \nonumber \\
&=& \sigma_{\max}(\mixMatrix)\|\Thetabf-\widehat \Thetabf \|_{F}.
\end{eqnarray}


Theorem \ref{ch5-th2} indicates $\delta^{*}_{\gamma' k}\leq 1/3$ as the sufficient condition for the sparse source recovery. In the following we investigate the implication of this condition for the previously mentioned acquisition schemes 
to bound the number of CS measurements.

\subsubsection{Dense Random Sampling}

Assume the compression matrix $A$ that is used for subsampling data in \eqref{sampling model2} is an $m \times \na\nb$ matrix whose elements are drawn independently at random from the Gaussian, Bernoulli or subgaussian distributions. According to Remark \ref{ch5-remark1}, such matrices satisfy D-RIP adapted to $\Phi$ (with the constant $\delta^{*}_{\gamma' k}\leq 1/3$) provided by:
\begin{eqnarray}\label{SS-meas dense}
m \gtrsim \gamma'k\,\log(\rho\na/\gamma'k)). 
\end{eqnarray}

\subsubsection{Uniform Random Sampling\label{sec_unif_rand_sampling}}

The same type of analysis indicates a very poor performance for the uniform random acquisition scheme described in section \ref{ch5-sec: Uniform sampling}. The corresponding sampling matrix has a block-diagonal form $A = \Id_{\nb}\otimes \atilde$. Here, we assume that the core compression matrix $\atilde$ that separately applies to each channel is an $\mhat \times \na$ matrix whose elements are drawn independently at random from Gaussian, Bernoulli or subgaussian distributions.


According to the theoretical analysis provided in section \ref{ch5-sec: RIP dic}, the sufficient condition for source recovery via \eqref{l1 SSCS} is $\delta_{k}(A) \leq \frac{ \sqrt 2-1-\delta_{k}(\Phi')} {1+\delta_{k}(\Phi')}$ which, by considering \eqref{ch5-eq2} can be rephrased as:
\[
\delta_{k}(A) \leq \frac{ \sqrt 2-1-\eta} {1+\eta}.
\]
For a compression matrix with this structure and by using Lemma \ref{RIP kron} we can deduce $\delta_{k}(A)\leq \delta_{k}(\atilde)$. Now similarly as for the bound \eqref{ch5-meas bound}, $\atilde$ satisfies the RIP with the constant above (and so does $A$) as long as $\mhat \geq c\, k\,\log(\na/k))$ or equivalently,
\begin{eqnarray}\label{SS-meas uniform}
m \geq c\, \nb\, k\,\log(\na/k)). 
\end{eqnarray}

The constant $c$ depends on the conditioning of the mixture matrix $\mixMatrix$. When the columns of $\mixMatrix$ are very coherent, the extreme singular values spread away from each other and $\eta$ becomes large. As a consequence, $\atilde$ (or equivalently $A$) must satisfy RIP for a smaller constant which, as  discussed earlier in section \ref{ch5-sec: RIP dic}, implies $c$ to be large and more CS measurements are required for  source recovery.  


\subsubsection{Decorrelating Random Sampling}


When a decorrelation step is incorporated into the compressive acquisition process, $\mixMatrix$ is discarded in the recovery formulation, and then  we can use the standard RIP analysis in \cite{donoho2006compressed, CRT-stable2005} to evaluate the source recovery performance.
Therefore, if $A=\Id_{\rho}\otimes \atilde$ satisfies the RIP with a constant $\delta_{k}(A)\leq \sqrt2-1$, then the solution $\widehat \Thetabf$ 
 to \eqref{l1 SSCS1} obeys the following error bound:
\begin{eqnarray}
 \| \Thetabf_{vec} - \widehat \Thetabf_{vec} \|_{2} \leq c_{0} \,k^{-1/2}  \|\Thetabf_{vec} - (\Thetabf_{vec})_{k}\|_{1} + c_{1} \varepsilon, \nonumber
 \end{eqnarray}
where the constants $c_{0}, c_{1}$ are the same as in \eqref{dic error bound}. 

Now, since $A$ is a block diagonal matrix, we can proceed along the exact same steps as for the uniform sampling scheme (Section \ref{sec_unif_rand_sampling}) to bound the minimum number of CS measurements such that $A$ satisfies the RIP:
\[
\mhat \geq \overline c\, k\, \log(\na/k).
\]
Unlike the previous measurement bounds for the non-decorrelating sampling schemes, here $\overline c$ is a fixed constant independent of the mixture matrix $\mixMatrix$. Consequently, the total number of CS measurements used for  source recovery is: 
\begin{eqnarray}\label{SS-meas decorr}
m \geq \overline c\, \rho\, k\,\log(\na/k)). 
\end{eqnarray}

Note that, for a noiseless sampling scenario ($\varepsilon=0$) the minimization \eqref{l1 SSCS1} can be decoupled into $\rho$ independent $\ell_{1}$ minimizations, each of them corresponding to a sparse recovery of a certain source. Now, if we assume that each source has exactly $ \k =k/\rho$ nonzero coefficients, then a perfect recovery can be guaranteed as long as $\delta_{ \k}(\atilde)\leq \sqrt2 -1$ which, for a matrix $\atilde$ drawn form the previously-mentioned distributions, implies that $\mhat \geq \overline c \, \k \log(\na/\k)$ and consequently:
\begin{eqnarray}\label{SS-meas decorr1}
m  = \rho \mhat \geq \overline c\, k  \log(\rho \na/k).
\end{eqnarray}

Comparing to \eqref{SS-meas decorr} where $m$ is roughly proportional to $\rho k$, here the measurement bound improves by a factor $\rho$ and it is mainly proportional to the sparsity level $k$ of all sources.

\subsection{Conclusions on the Theoretical Bounds}\label{ch5-conclusion1}
Consider a multichannel data derived by the linear mixture \eqref{mix model} of $\rho$ sources, each having a $k'$-sparse representation \ie $\sbf$ is $k=\rho k'$ sparse. 
Table \ref{table:count1} summarizes the scaling-orders of the number of CS measurements sufficient for an exact data reconstruction for different noiseless random acquisition schemes and sparse recovery approaches. 
\begin{table*}[t!]
\small
\begin{center}
\resizebox{\linewidth}{!}{
\begin{tabular}{|c|c|c|c|c|}
\hline
\textbf{CS Acquisition Scheme} & Dense  & Dense & Uniform & Decorrelating \\
\hline
\textbf{CS Recovery Approach} & BPDN &  SS-$\ell_{1}$ & SS-$\ell_{1}$ & SS-$\ell_{1}$\\
\hline
CS measurements $m \gtrsim $&  $\O\Big(\nb k \log(\na/k) \Big)$  &  $\O\Big( k\log(\rho\na/k) \Big)$ & $\O \Big( \nb k\log(\na/k) \Big)$ &$\O\Big( k\log(\rho\na/k) \Big)$   \\
\hline
Constant depends on $\mixMatrix$ &  -   & Yes & Yes & No  \\
\hline
\end{tabular}
}
\end{center}
\caption{Measurement bounds for random sampling schemes: dense, uniform and decorrelating,  
and for recovery approaches: BPDN and SS-$\ell_{1}$ (\ie source separation based recovery using \eqref{l1 SSCS} or \eqref{l1 SSCS1}). The last row shows if the bounds for the SS-$\ell_{1}$ are sensitive to the conditioning of the mixing matrix $\mixMatrix$.
}\label{table:count1}
\end{table*}
As we can observe, compressed sensing via source recovery using \eqref{l1 SSCS} once it is coupled with a proper CS acquisition (\ie, Dense i.i.d. subgaussian $A$, or a random decorrelating sampling scheme as in sections \ref{ch5-sec: dense decorr sampling} and \ref{ch5-sec: uniform decorr sampling}) leads to a significantly improved  bound compared to standard methods such as BPDN. 
More remarkably, the number of CS measurements turns out to be independent of the number $\nb$ of channels.


Finally note that the measurement bound for the source-separation-based reconstruction approach, which uses a non-decorrelating random compression matrix, depends on the conditioning of the mixture parameters via the constant factor $\gamma'$ in \eqref{SS-meas dense}. Therefore, when the columns of $\mixMatrix$ are highly coherent, the condition number of $\mixMatrix$ becomes relatively large,  and so does $\gamma'$.
This limitation can be circumvented thanks to the decorrelating acquisition scheme.

\section{Applications in Compressive Hyperspectral Imagery}\label{ch5-app}

Compressed sensing is particularly promising for  hyperspectral imagery where  the acquisition procedure is very costly.
This type of images can be approximated by a linear mixture model as in \eqref{mix model}
where each spatial pixel is populated with a very few number of materials (i.e. sources). 
 In this regard, $\sbf \in [0,1]^{\na \times \rho} $ is a matrix whose $\rho$ columns are  \emph{source images} (vectorized 2D images) indicating the percentage of each material in one of the $\na$ spatial pixels, and therefore
 \begin{equation}\label{rowsum1}
\sum_{j=1}^{\rho} [\sbf]_{i,j}=1 \qquad \forall i \in \{1,\ldots,\na\}.
\end{equation}
Moreover, $\mixMatrix \in \Rbb_{+}^{\nb\times \rho}$ is a matrix whose columns contain the spectral signatures of the corresponding sources of $\sbf$. Note that in some particular applications and specially when the spatial resolution is high enough, the source images become \emph{disjoint}, meaning that each spatial pixel contains only one material and $[\sbf]_{i,j} \in \{0,1\}$.

The two key priors that will be essential for  compressive source identification are the following: i) Each source image contains piecewise smooth variations along the spatial domain, implying a sparse representation in a wavelet basis, or sparsity of its gradient, and ii) each spatial pixel is a non-negative linear combination of a \emph{small} number of sources.

In the next two sections we introduce two classes of source separation based recovery approaches that are particularly adapted to  hyperspectral compressive imagery. 

\subsection{Compressive HSI Source Separation via Convex Minimization}
According to our earlier assumptions, source images are spatially piecewise smooth, which means the coefficients $\Thetabf$
of $\sbf = \Psia \Thetabf$ are sparse in a 2-dimensional wavelet basis $\Psia \in \Rbb^{\na\times \na}$.
We conveniently rephrase this representation  in  a vectorized form $\sbf_{vec} = \Psibf \Thetabf_{vec}$
with $\Psibf = \Id_{\rho}\otimes \Psia$ as described in Section \ref{sec_sparse_reg_MultArra_sig}.

Taking into account the sparsity of $\Thetabf_{vec}$ and by incorporating specific assumptions such as \eqref{rowsum1} and non-negativity we can extend the $\ell_{1}$ minimization approach in \eqref{l1 SSCS} as follows:
\begin{align}\label{HSI l1 SS}
\argmin_{\Thetabf} \quad & \| \Thetabf_{vec}\|_{1}    \\
\text{subject to}\quad  & \left\| y - A \Phi \Psibf \Thetabf_{vec} \right\|_{2} \leq \varepsilon \nonumber\\
& \Psia\, \Thetabf\, \u_{\rho} = \u_{\na}\nonumber \\
& \Psibf \Thetabf_{vec}\geq 0. \nonumber
\end{align}
Where, $\u_{n}$ denotes an all one $n$-dimensional vector. The first constraint is the same as the fidelity constraint in \eqref{l1 SSCS}. 
The last two constraints impose the element-wise non-negativity of $\sbf$ and the ``percentage'' normalization \eqref{rowsum1} \ie, each row of $\sbf$ belongs to the positive face of the simplex in $\Rbb^{\rho}$. 
Minimizing the  $\ell_{1}$ norm together with the last two constraints (that is  equivalent to an additional $\ell_{1}$ norm constraint) gives solutions that contain both desired sorts of sparsity: i) along the 2D wavelet coefficients of $\sbf$ and, ii) along each row of $\sbf$. 

Note that the theoretical analysis given in  Section \ref{ch5-SStheory} can also apply here to bound the performance of \eqref{HSI l1 SS}.  
Although we bound the error similarly as for \eqref{l1 SSCS}, one can naturally expect a much better performance  for \eqref{HSI l1 SS}
thanks to the two additional constrains.

Alternatively, problem \eqref{HSI l1 SS} can be formulated in a more general ``analysis'' formulation with an analysis sparsity prior $\sparsPrior(\sbf)$:
\begin{align}\label{HSI Analysis SS}
\argmin_{\sbf} \quad & \sparsPrior(\sbf)    \\
\text{subject to}\quad  & \left\| y - A \Phi \sbf_{vec} \right\|_{2} \leq \varepsilon \nonumber\\
& \sbf\, \u_{\rho} = \u_{\na}\nonumber \\
& \sbf_{vec}\geq 0. \nonumber
\end{align}
which is equivalent to \eqref{HSI l1 SS}  when $\sparsPrior(\sbf) = \|\Psibf^{*} \sbf_{vec}\|_{1}$ and $\Psibf$ is a square and invertible operator. Another efficient analysis prior for image regularization is the Total Variation which can be applied on each source image of the HSI  with the prior: $\sparsPrior(\sbf) = \sum_{j}\|\sbf_{j}\|_{TV}$.
The problem formulation \eqref{HSI Analysis SS} is general and includes the decorrelating schemes discussed in sections \ref{ch5-sec: dense decorr sampling} and \ref{ch5-sec: uniform decorr sampling}. Indeed inserting the  matrix $A$ of \eqref{decorr matrix} in  \eqref{HSI Analysis SS} leads to  the following fidelity term $\|y - \usm[\rho] \,\sbf_{vec} \|_{2} \leq \varepsilon$ while the other terms remain unchanged. 



In the next Section we provide an iterative algorithm for solving problem \eqref{HSI Analysis SS}. 
When sources are disjoint, it is also possible to add a \emph{hard thresholding} post-processing step that sets the maximum coefficient of each row of $\widehat \sbf$ equal to one and set to zero the other coefficients.

\subsection{The PPXA Algorithm for Compressive Source Separation}

The Parallel Proximal Splitting Algorithm (PPXA) \cite{proximal-splitting} is an iterative method for minimizing an arbitrarily finite sum of  lower semi-continuous (l.s.c.) convex  functions. Each of the iteration consists in computing the \emph{proximity} operator of all functions (which can be done in parallel), averaging their results and updating the solution until convergence. The proximity operator of a function $f(x) :\Rbb^{n} \rightarrow \Rbb$ is defined   as $\prox_{f}: \Rbb^{n}\rightarrow  \Rbb^{n}$\cite{proximal-splitting}: 
\begin{eqnarray}\label{ch5-prox}
\underset{\widetilde x \in  \Rbb^{n}} {\argmin} f (\widetilde x) + \frac{1}{2}\|x - \widetilde x \|_{2}^{2}.
\end{eqnarray}

For solving \eqref{HSI Analysis SS} with PPXA, we rewrite it as the minimization of the sum of three l.s.c. convex functions:
\begin{eqnarray}\label{ch5-sumPPXA}
\argmin_{\sbf} f_{1}(\sbf)+f_{2}(\sbf)+f_{3}(\sbf),
\end{eqnarray}
with $f_{1}(\sbf)  = \sparsPrior(\sbf)$, 
$f_{2}(\sbf) = i_{\mathcal B_{2}}(\sbf)$ and  $f_{3}(\sbf) = i_{\mathcal B_{\Delta+}}(\sbf)$ and where
 $i_{\mathcal{C}}$ is the indicator function of a convex set $\mathcal{C}$ defined as:
\begin{equation}
i_{\mathcal C}(\sbf) = \left\{ 
  \begin{array}{ll}
    0 & \quad \text{if}\,\,\sbf \in {\mathcal C} \nonumber\\
    +\infty &  \quad \text{otherwise},\\
  \end{array} \right.
\end{equation}
\noindent and the convex sets $\mathcal B_{2}, \mathcal B_{\Delta+} \subset \Rbb^{\na \times \rho}$ are respectively, the set of matrices that  satisfy the fidelity constraint $\|y-A\Phi\sbf_{vec}\|_{2} \leq \varepsilon$, and the set of matrices whose rows belong to the standard simplex in $\Rbb^{\rho}$.
The template of the PPXA algorithm that solves \eqref{ch5-sumPPXA} and hence \eqref{HSI Analysis SS} is given in Algorithm 1. 
We now derive the proximity operator of each function $f_{i}$.
Note that the definition of the proximity operator in \eqref{ch5-prox}  naturally extends for matrices by replacing the $\ell_2$ norm with the Frobenius norm.


\begin{algorithm}  [t!]        \label{ch5-PPXA}           
\SetAlgoLined \KwIn{ $y$, $A$, $\Phi$, $\varepsilon$, $\beta > 0$.} 
\textbf{Initializations:} \\
\vspace{-.2cm}
$n=0$, $\sbf_{0}=\Gamma_{1,0}=\Gamma_{2,0}=\Gamma_{3,0} \in \Rbb^{n_{1}\times n_{2}}$\\
\vspace{-.2cm}
\Repeat
{convergence}{
\For{$(i =1: 3)$ }
{$P_{i,n} = \prox_{3\beta f_{i} }(\Gamma_{i,n})$\\}
$\sbf_{n+1} =(P_{1,n}+P_{2,n}+P_{3,n})/3$\\ 
\For{$(i =1: 3)$ }
{$\Gamma_{i,n+1} = \Gamma_{i,n} + 2\sbf_{n+1} - \sbf_{n} - P_{i,n}$\\}
}
\caption{The Parallel Proximal Algorithm to solve \eqref{HSI Analysis SS}. } 
\end{algorithm}

For $\sparsPrior(\sbf) = \|\Psibf^{*} \sbf_{vec}\|_{1}$, a standard calculation shows that 
\begin{equation}
(\prox_{\alpha{\sparsPrior}})_{i} = \text{sign}\big((\Psibf^{*}\sbf_{vec})_{i}\big)\,.\, \big( | (\Psibf^{*}\sbf_{vec})_{i} | - \alpha \big)_{+},
\end{equation}
which is the \emph{soft thresholding} operator applied on the wavelet coefficients of $\sbf$.  
The  proximity operator of $\sparsPrior(\sbf) = \sum_{j=1}^{\rho} \| \sbf_{j} \|_{TV}$ can be decoupled and computed in parallel for each of the $\rho $ sources via 
an efficient implementation proposed by \cite{TV-chambolle}. 
By definition, the proximal operator of an indicator function $i_{\mathcal C}(\sbf)$ is the orthogonal projection of $\sbf$ onto the corresponding set $\mathcal C$. 
The projection onto the standard simplex ${\mathcal{B}_{\Delta+}}$ can be done in one iteration using the method proposed by Duchi et al. \cite{Duchi}. For a general implicit operator $\opAPsi \defeq A\Phi$, the projector onto ${\mathcal{B}_{2}}$ can be computed
using a \textit{forward backward} scheme as proposed in \cite{FadiliS09}.
This projection usually has the dominant computational complexity of the algorithm because of costly sub-iterations.
However if the decorrelating sampling scheme is used and $\opAPsi = \usm[\rho]$ is a tight frame (\ie, $\forall x\in \Rbb^{\mhat}$ $\opAPsi \opAPsi^{*}x = \nu\, x$ for a constant $\nu$), then according to the \textit{semi-orthogonal linear transform} property of proximity operators \cite{proximal-splitting},  the orthogonal projection onto ${\mathcal{B}_{2}}$ has the following explicit form:
\vspace{-.2cm}
 \begin{eqnarray}
\left( \textit{prox}_{\alpha f_{2}}(\sbf)\right)_{vec}  = 
    \sbf_{vec} + \frac{1}{\nu} (\usm[\rho])^{*} \residdecorMtx
    \left(1- \frac{\varepsilon}{  \left\| \residdecorMtx \right\|_{2} } \right)_{+},\label{eq:closed_form_prox_f2}
\end{eqnarray}
with $\residdecorMtx = y -\usm[\rho] \sbf_{vec}$.

\subsection{Compressive HSI Source Separation via Iterative Hard Thresholding}

If the source images are disjoint, 
the following non-convex minimization can be alternatively used for recovering the sparse wavelet coefficients of the sources:
\begin{align}\label{IHT SS}
\argmin_{\Thetabf} \quad & \left\| y - A \Phi \Psibf \Thetabf_{vec} \right\|^{2}_{2} \\
 \text{subject to}\quad  & \| \Thetabf_{vec} \|_0 \leq k \nonumber\\
 & \offdiag(\Thetabf^{*} \Thetabf)=0 \nonumber\\
& \Psia\, \Thetabf\, \u_{\rho} = \u_{\na}\nonumber \\
& \Psibf \Thetabf_{vec}\geq 0. \nonumber
\end{align}
where the operator $\offdiag(B)$ returns the off-diagonal elements of matrix  $B$, and 
the $\ell_{0}$ norm constraint on $\Thetabf_{vec}$ imposes the wavelet coefficients to be $k$-sparse.
The second constraint imposes the orthogonality of the wavelet coefficients which is a consequence of the source disjointness.
The two last constraints are the same as in \eqref{HSI l1 SS}.

\begin{algorithm}  [t]        \label{alg:IHT SS}           
\SetAlgoLined \KwIn{ $y$, $A$, $\Phi$,   $\gamma = 1/\|A \Phi \Psibf \|^{2} = 1/\|A \Phi \|^{2}$ and  $k$.} 
\textbf{Initializations:} \\
$n=0$, $\Thetabf^{0} \in \Rbb^{\na\times \rho}$\\

\Repeat
{convergence}{
1- Gradient descent:\\
$\qquad\Thetabf^{n+1}_{vec} = \Thetabf^{n}_{vec} - \gamma \nabla F(\Thetabf^{n})$\\
2- Hard thresholding: \\
$\qquad\Thetabf^{n+1}_{vec}  = \text{Th}_{k}( \Thetabf^{n+1}_{vec})$\\
3- Orthogonal matrix procrustes:\\
$\qquad \text{Update}\,\, \Omega:$\,\,$[\Omega]_{i,i} = \sqrt{\na} \,\frac{ \|\Thetabf^{n+1}_{.,i}\|_{2} }{\|\Thetabf^{n+1}\|_{F}}$\\
$\qquad \text{Singular value decomposition:} \,\,U \Sigma V^{*} = \Thetabf^{n+1}\Omega$\\
$\qquad \Thetabf^{n+1} = UV^{*}\Omega$\\
4- Simplex projection:\\
$\qquad \Thetabf^{n+1} = \Psia^{*}\,\, \text{Project}_{\mathcal B_{\Delta+}} (\Psia \Thetabf^{n+1})$
}
\caption{The Iterative Hard Thresholding Algorithm to approximate solution of \eqref{IHT SS}} 
\end{algorithm}

Despite its convex objective term, \eqref{IHT SS} has multiple non-convex constraints and is therefore a non-convex problem. 
We propose an algorithm similar to the \emph{Iterative Hard Thresholding} (IHT) algorithm \cite{IHT-sparse}
to approximate the solution of \eqref{IHT SS}. 
At each iteration the current solution is updated by a gradient descent step followed by a hard thresholding step $\text{Th}_{k}(\cdot)$ that selects the $k$ largest wavelet coefficients of $\widehat \Thetabf_{vec}$. In addition the three last constraints of \eqref{IHT SS} are applied sequentially: 
\begin{itemize}
\item First, a procedure inspired by the \emph{orthogonal matrix procrustes} is applied to diagonalize $\widehat \Thetabf^{*} \widehat \Thetabf$. Let $\Omega$ be a $\rho \times \rho$ diagonal matrix where for $1\leq i \leq \rho$ we have
\[[\Omega]_{i,i} = \sqrt{\na} \,\frac{ \|\widehat \Thetabf_{.,i}\|_{2} }{\|\widehat \Thetabf\|_{F}}.\]
Since for disjoint sources we have $\|\sbf\|_{F}=\|\Thetabf\|_{F}=\sqrt \na$, then a good orthogonal matrix that would approximate $\widehat \Thetabf$ and keeps the energy of the current estimate of each source image proportional to that of the previous estimate 
would be $UV^{*}\Omega$ 
 through the following singular value decomposition $U \Sigma V^{*}= \widehat \Thetabf \Omega$.

\item Second, the current solution $\widehat \sbf = \Psia \widehat \Thetabf$ is projected onto the standard simplex  as in \cite{Duchi}. 
\end{itemize}

The description of the this algorithm can be found in Algorithm 2. 
Note that the gradient of the objective functional $F (\Thetabf) = \left\| y - A \Phi \Psibf \Thetabf_{vec} \right\|^{2}_{2}$  is:
\begin{equation}
\label{eq:grad_dense}
\nabla F(\Thetabf) = - (A \Phi \Psibf)^{*} \big( y - A \Phi \Psibf \Thetabf_{vec} \big).
\end{equation}
Using the decorrelating scheme, the objective function
in \eqref{IHT SS} becomes $F(\Thetabf) = \| y - \atilde_{\rho} \Psibf \Thetabf_{vec} \|^{2}_{2}$ with gradient :
\begin{equation}
\nabla F(\Thetabf) = - (\atilde_{\rho} \Psibf)^{*} \big( y - \atilde_{\rho} \Psibf \Thetabf_{vec} \big).\label{eq:grad_decor}
\end{equation}
The rest of Algorithm 2 
remains unchanged.

In the next section, we evaluate the performances of these algorithms on HSI. 

\section{Experiments \label{sec_expe}}

In this section, we evaluate the ability of the methods presented in Section \ref{ch5-app}, 
(called ``\SSMethods'' and summed up in table \ref{tab:methods}) 
 to separate the sources and recover HSI in various scenarios: 
 various noise levels (from  noiseless  to $10$ dB SNR), various sampling ratios (from ${m}/{(\na\nb)} = 1/4$ to $1/32$ sampling rates), various sampling mechanisms (uniform and dense sampling), on two different HSI (Geneva and Urban).
We also compare the \SSMethods\ with the classical methods for CS,
such as the 
BPDN problem \eqref{eq_BPDN_NA_Basis} \solveBPDNone, the TVDN problem \eqref{eq_BPDN_NA_TV} \solveTVDN,
both solved with a Douglas-Rachford (DR) splitting algorithm.
%
\begin{table*}
\caption{Description of the proposed \SSMethods.}
\label{tab:methods}
\centering
\begin{tabular}{|c|>{\footnotesize}c|>{\footnotesize}c|}
  \hline
  Method name & \multicolumn{1}{|c|}{Description} \\
    \hline
  \fastSSIHTfinal & 
   \begin{minipage}{\descriptionwidth}
   \vspace{\tabvspace}
  Problem \eqref{IHT SS} solved with Algorithm 2 with gradient $\nabla F(\Thetabf)$ of Eq. \eqref{eq:grad_dense}.
   \vspace{\tabvspace}
   \end{minipage}\\   
    \hline
   \fastSSlonefinal & 
      \begin{minipage}{\descriptionwidth}
   \vspace{\tabvspace}
   Problem \eqref{HSI Analysis SS} solved with Algorithm 1, with  $\sparsPrior(\sbf) = \|\Psibf^{*} \sbf_{vec}\|_{1}$
   and $\prox_{\alpha f_{2}}(\cdot)$ computed using a forward-backward scheme as proposed in \cite{FadiliS09}.
      \vspace{\tabvspace}
   \end{minipage}
   \\
     \hline
   \fastSSTVfinal & 
         \begin{minipage}{\descriptionwidth}
   \vspace{\tabvspace}
   Problem \eqref{HSI Analysis SS} solved with Algorithm 1, with  $\sparsPrior(\sbf) = \sum_{j=1}^{\rho} \| \sbf_{j} \|_{TV}$ 
   and $\prox_{\alpha f_{2}}(\cdot)$ computed using a forward-backward scheme as proposed in \cite{FadiliS09}.
   \vspace{\tabvspace}
   \end{minipage}
   \\
     \hline
   \fastSSstIHTfinalProj & 
   \begin{minipage}{\descriptionwidth}
   \vspace{\tabvspace}
   Problem \eqref{IHT SS} solved with Algorithm 2 with gradient $\nabla F(\Thetabf)$ of Eq. \eqref{eq:grad_decor}.
   \vspace{\tabvspace}
   \end{minipage}\\   
     \hline
   \fastSSlonefinalProj & 
    \begin{minipage}{\descriptionwidth}
   \vspace{\tabvspace}
   Problem \eqref{HSI Analysis SS} solved with Algorithm 1, with  $\sparsPrior(\sbf) = \|\Psibf^{*} \sbf_{vec}\|_{1}$, 
   and $\prox_{\alpha f_{2}}(\cdot)$ computed with the closed form Eq. \eqref{eq:closed_form_prox_f2}.
   \vspace{\tabvspace}
   \end{minipage}
   \\
     \hline
   \fastSSTVfinalProj & 
   \begin{minipage}{\descriptionwidth}
   \vspace{\tabvspace}
   Problem \eqref{HSI Analysis SS} solved with Algorithm 1, with  $\sparsPrior(\sbf) = \sum_{j=1}^{\rho} \| \sbf_{j} \|_{TV}$ and $\prox_{\alpha f_{2}}(\cdot)$ computed with \eqref{eq:closed_form_prox_f2}.
    \vspace{\tabvspace}
   \end{minipage}
   \\
  \hline
\end{tabular}
\end{table*}

\subsection{Sampling Mechanism}

We used two different sampling schemes: i) the sensing matrix $A$ is \textit{dense} (and the methods implementing the decorrelation step cannot be applied), 
and ii) \textit{uniform} sampling where the sensing matrix is block diagonal with identical blocks as in \eqref{eq:uniform sensing matrix}.
In the latter, the decorrelation step can be applied as explained in section \ref{sec_decorr_sect}.

So as to generate the random sampling matrices $A$ and $\atilde$ that can be used in practical applications, 
we used the Random Convolution (RC) measurement scheme proposed by Romberg \cite{RCromberg} that convolves the image with a random pattern using few optical blocks. More remarkably, sampling matrices generated by RC are tight frames and thus for decorrelating schemes, they benefit from a closed form expression \eqref{eq:closed_form_prox_f2} for computing $\prox_{\alpha f_{2}}(\cdot)$  that can massively accelerates the recovery procedure.
\subsection{The Geneva HSI \label{sec:GENEVA_HSI}}

\if@twocolumn
\else
\begin{figure}[t]
    \centering \subfigure[Reconstruction SNR vs. subsampling ratio (noiseless sampling)]{
    \label{fig:snr vs meas}
    \begin{minipage}[b]{.46\linewidth}
      \centering \includegraphics[width=.95\linewidth]{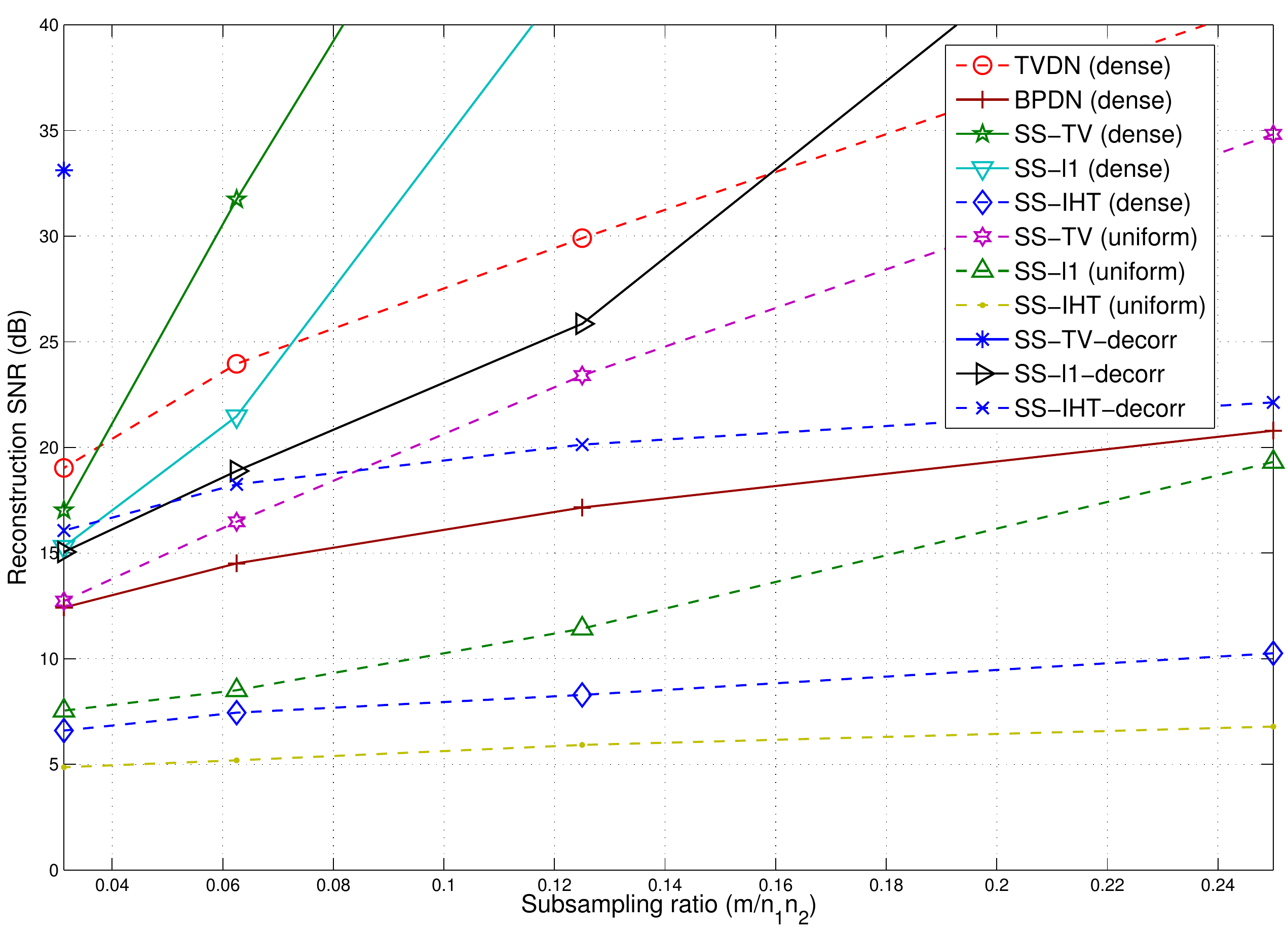}
    \end{minipage}}%
    \hfil
        \centering \subfigure[Reconstruction SNR vs. sampling SNR (subsampling ratio:$1/16$)]{
    \label{fig:snr vs noise}
    \begin{minipage}[b]{.46\linewidth}
      \centering \includegraphics[width= .95\linewidth]{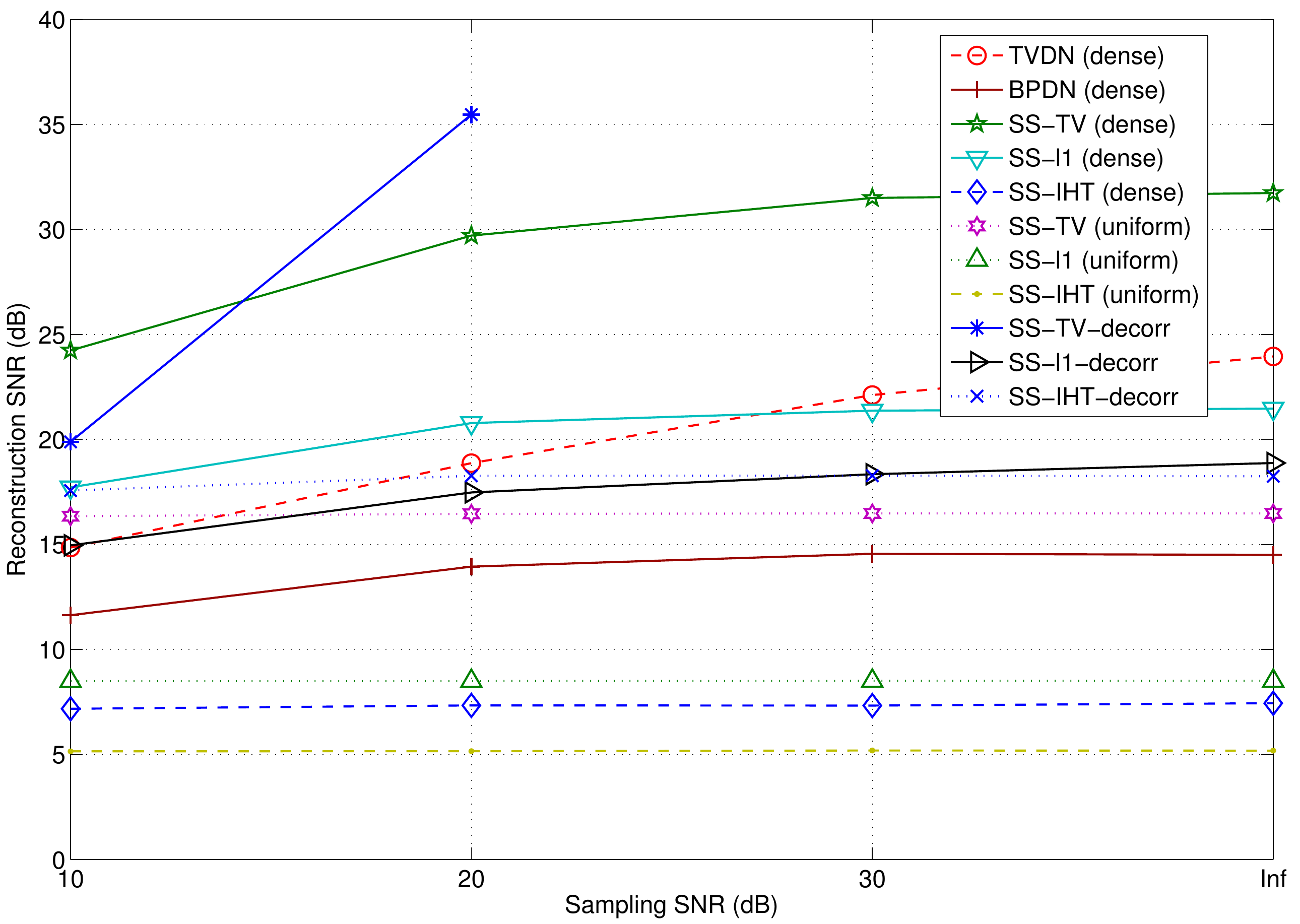}
    \end{minipage}}%
      \caption{Geneva HSI reconstruction performance for different sampling mechanisms and recovery methods. Points with $\infty$ reconstruction SNR (exact recovery) are not plotted. \label{fig:GVA rec}}
\end{figure}
\fi

We evaluate the different methods, 
for different sampling rates (Fig. \ref{fig:snr vs meas}), and different noise levels (Fig. \ref{fig:snr vs noise}), 
on a HSI generated from a ground truth map image \footnote{We acknowledge Xavier Gigandet and Meritxell Bach Cuadra for providing this ground truth map.} 
of farms in a suburb of Geneva. The source spectra (i.e. columns of $\mixMatrix$)
 are chosen form the USGS digital spectral library \cite{speclabUSGS}.
The HSI cube has spatial slices of the resolution $N=256 \times 256$ that are taken over $J=224$ frequency bands. 

%
%
%



\subsubsection{Performance of the \SSMethods}



\begin{table*}
\caption{Source separation performance (Accuracy) of \SSMethods. Methods with the highest accuracy are highlighted in each column.}
\label{tab:SS_performance}
\centering
\begin{tabular}{|c|c|c|c|c|c|c|c|c|c|c|c|c|}
  \hline  Noise SNR & \multicolumn{4}{c|}{ $+\infty$ dB } &  \multicolumn{4}{c|}{ $30$ dB } &  \multicolumn{4}{c|}{ $10$ dB }\\
  \hline Sampling rate & $1/4$ & $1/8$ & $1/16$ & $1/32$ & $1/4$ & $1/8$ & $1/16$ & $1/32$ & $1/4$ & $1/8$ & $1/16$ & $1/32$\\
  \hline 
  \hline \fastSSIHTfinal \textit{(dense sampling)} & 0.69 & 0.61 & 0.57 & 0.48 &  0.71 & 0.6 & 0.57 & 0.48 & 0.7 &  0.6 & 0.57 & 0.48\\
  \hline \fastSSlonefinal  \textit{(dense sampling)} & \textbf{1.0} &  \textbf{1.0} & 0.95 & 0.81 & \textbf{1.0} & \textbf{1.0} & 0.95 & 0.8 & \textbf{1.0} & 0.98 & 0.91 & 0.73\\
  \hline \fastSSTVfinal  \textit{(dense sampling)} & \textbf{1.0} &  \textbf{1.0} &  \textbf{1.0} & 0.92 &  \textbf{1.0} & \textbf{1.0} &  \textbf{1.0} & 0.91 & \textbf{1.0} &  \textbf{1.0} & \textbf{0.98} & 0.88\\
  
  \hline \fastSSIHTfinal  \textit{(uniform sampling)}  & 0.43 & 0.38 & 0.31& 0.25 & 0.43 & 0.37 & 0.31 & 0.26 & 0.43 & 0.37 &  0.3 & 0.26\\
  \hline \fastSSlonefinal \textit{(uniform sampling)}&  0.97 & 0.73 & 0.45 & 0.31 & 0.95 & 0.73 & 0.48 & 0.3 & 0.96 & 0.75 & 0.42 & 0.3\\
  \hline \fastSSTVfinal \textit{(uniform sampling)}& \textbf{1.0} & 0.98 & 0.9 & 0.76 &  \textbf{1.0} & 0.97 & 0.89 & 0.74 & \textbf{1.0} & 0.97 & 0.88 & 0.74\\
  
  \hline \fastSSstIHTfinalProj & 0.98  &   0.98 &  0.96 &  0.94 & 0.99 & 0.98 & 0.96 & 0.94 &  0.98 & 0.97 & 0.95 & 0.92\\

  \hline \fastSSlonefinalProj &\textbf{1.0}  &  0.99  &  0.97 &  0.92 & \textbf{1.0} & 0.99 & 0.96 & 0.91& 0.98 & 0.95 & 0.92 & 0.87\\
  \hline \fastSSTVfinalProj & \textbf{1.0}  &  \textbf{1.0}  & \textbf{ 1.0} &  \textbf{1.0} & \textbf{1.0} &  \textbf{1.0} &  \textbf{1.0} & \textbf{1.0} & \textbf{1.0} & 0.99 & \textbf{0.98} & \textbf{0.96}\\
  \hline
\end{tabular}
\end{table*}

\if@twocolumn

 \else
 
 \begin{figure}[ht]

\centering  \subfigure[True sources \label{fig:src_Geneva}]{ \label{}
 \includegraphics[width= .7\textwidth]{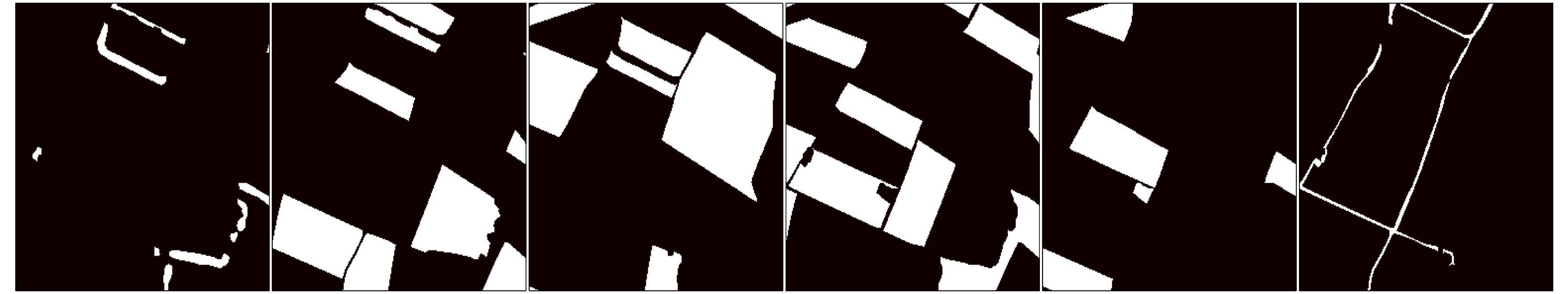}}

\centering \subfigure[\fastSSTVfinal\  \emph{(dense sampling)}  ]{ \label{}
\includegraphics[width= .7\textwidth]{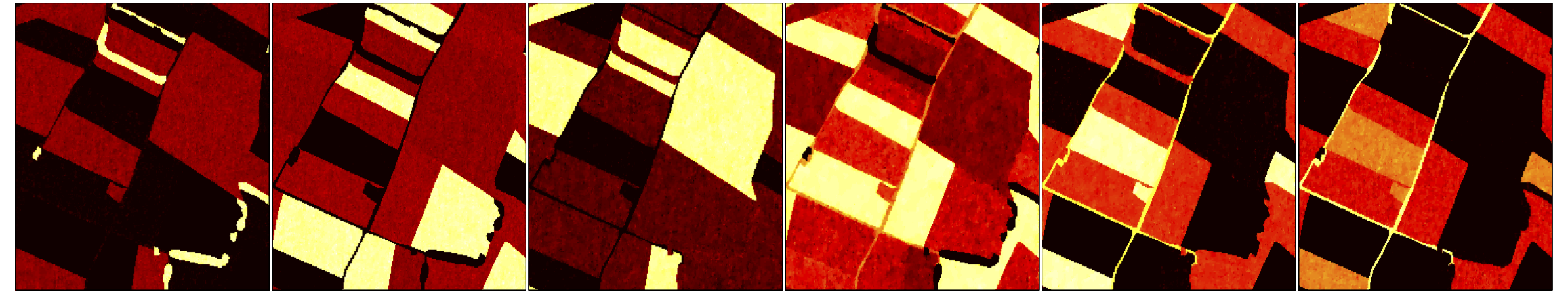} }

   \centering   \subfigure[\fastSSTVfinal\  \emph{(uniform non-decorrelating sampling)}      ]{ \label{}
    
 \centering    \includegraphics[width=.7\textwidth]{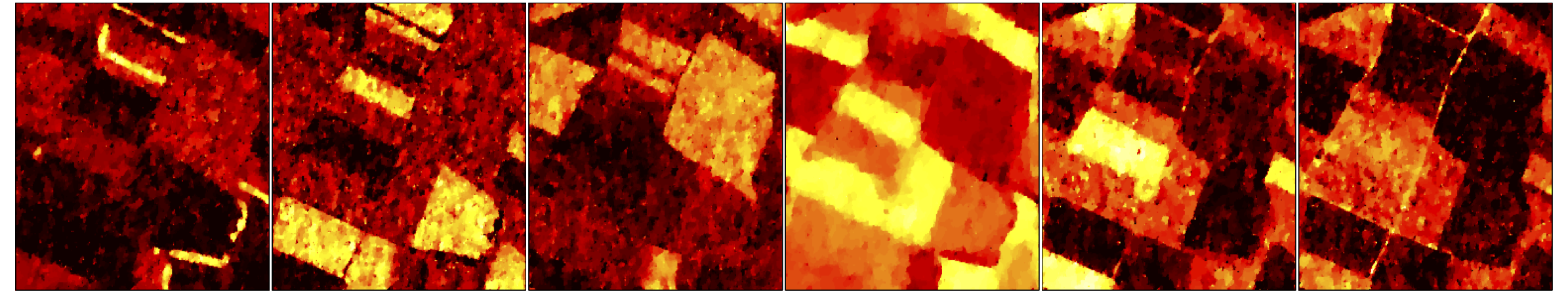} }

\centering     \subfigure[\fastSSTVfinalProj\   \emph{(uniform decorrelating sampling)}     ]{ \label{}
\centering \includegraphics[width= .7\textwidth]{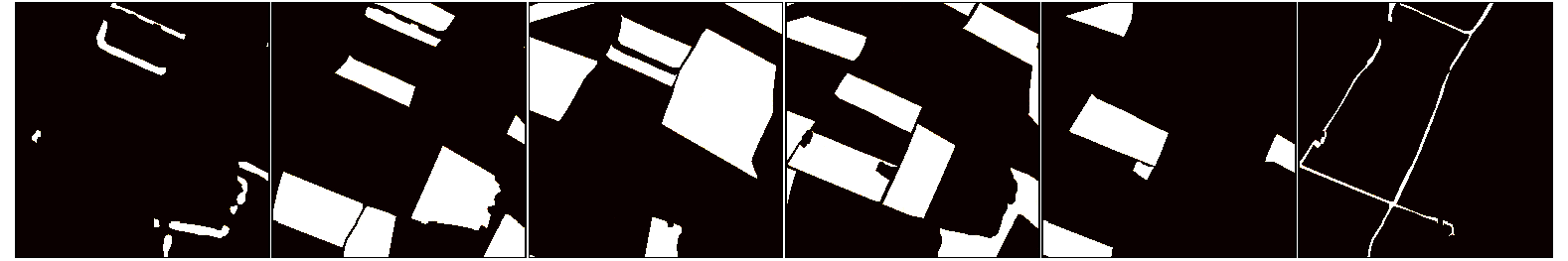}}%

\centering \subfigure[\fastSSstIHTfinalProj\  \emph{(uniform decorrelating sampling)}]{ \label{}
 
\centering   \includegraphics[width=.7 \textwidth]{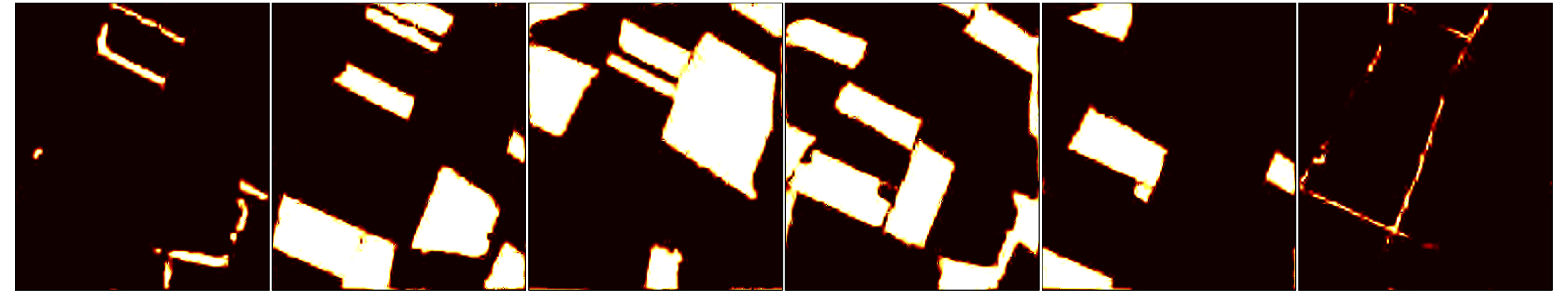}}
    
  \caption{Estimated source images of Geneva HSI for different sampling schemes and recovery methods (subsampling ratio: 1/16, noiseless sampling). \label{fig:rec_sources_Geneva}}

\end{figure}
\fi


Concerning the performance of the \SSMethods, we observe in Fig. \ref{fig:GVA rec} that: 
\begin{itemize}
\item The dense sampling scheme is always better than the uniform sampling scheme.
\item The decorrelated scheme is always better than dense sampling for $TV$-based and IHT methods, but is not for the $\ell_1$-based method.
\item  The decorrelating method \fastSSTVfinalProj\ results in perfect reconstruction 
in the cases where the sampling ratio is higher or equal to $1/16$ and performs better than all the other methods in all  regimes, except in high noise of $10$ dB SNR, where the dense approach \fastSSTVfinal\  (\textit{dense sampling}) performs slightly better.
\end{itemize}

\if@twocolumn
\begin{figure}[t]
    \centering \subfigure[Reconstruction SNR vs. subsampling ratio (noiseless sampling)]{
    \label{fig:snr vs meas}
    \begin{minipage}[b]{\linewidth}
      \centering \includegraphics[width= .9\textwidth]{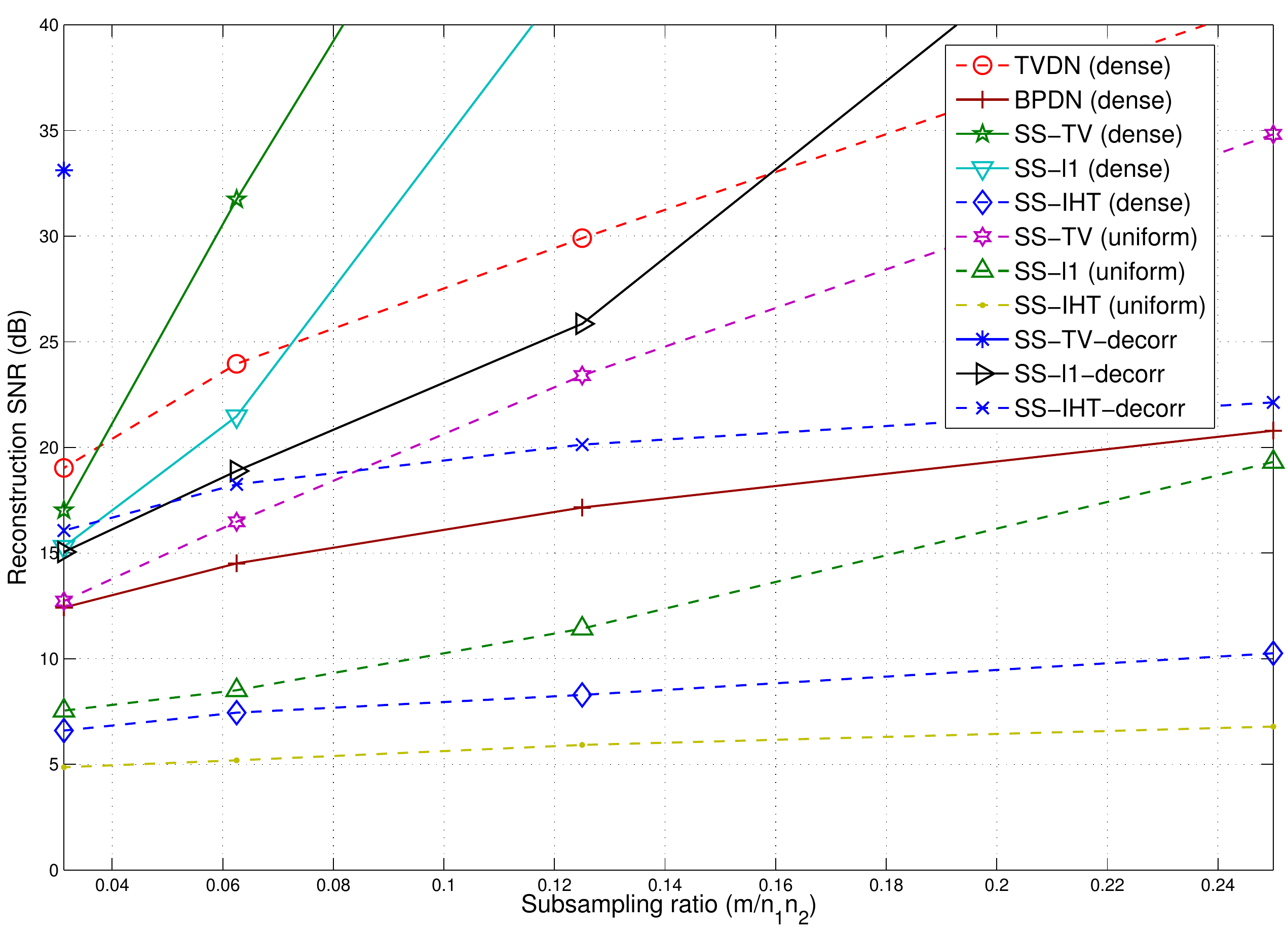}
    \end{minipage}}%
    \hfil
        \centering \subfigure[Reconstruction SNR vs. sampling SNR (subsampling ratio:$1/16$)]{
    \label{fig:snr vs noise}
    \begin{minipage}[b]{\linewidth}
      \centering \includegraphics[width= .9\textwidth]{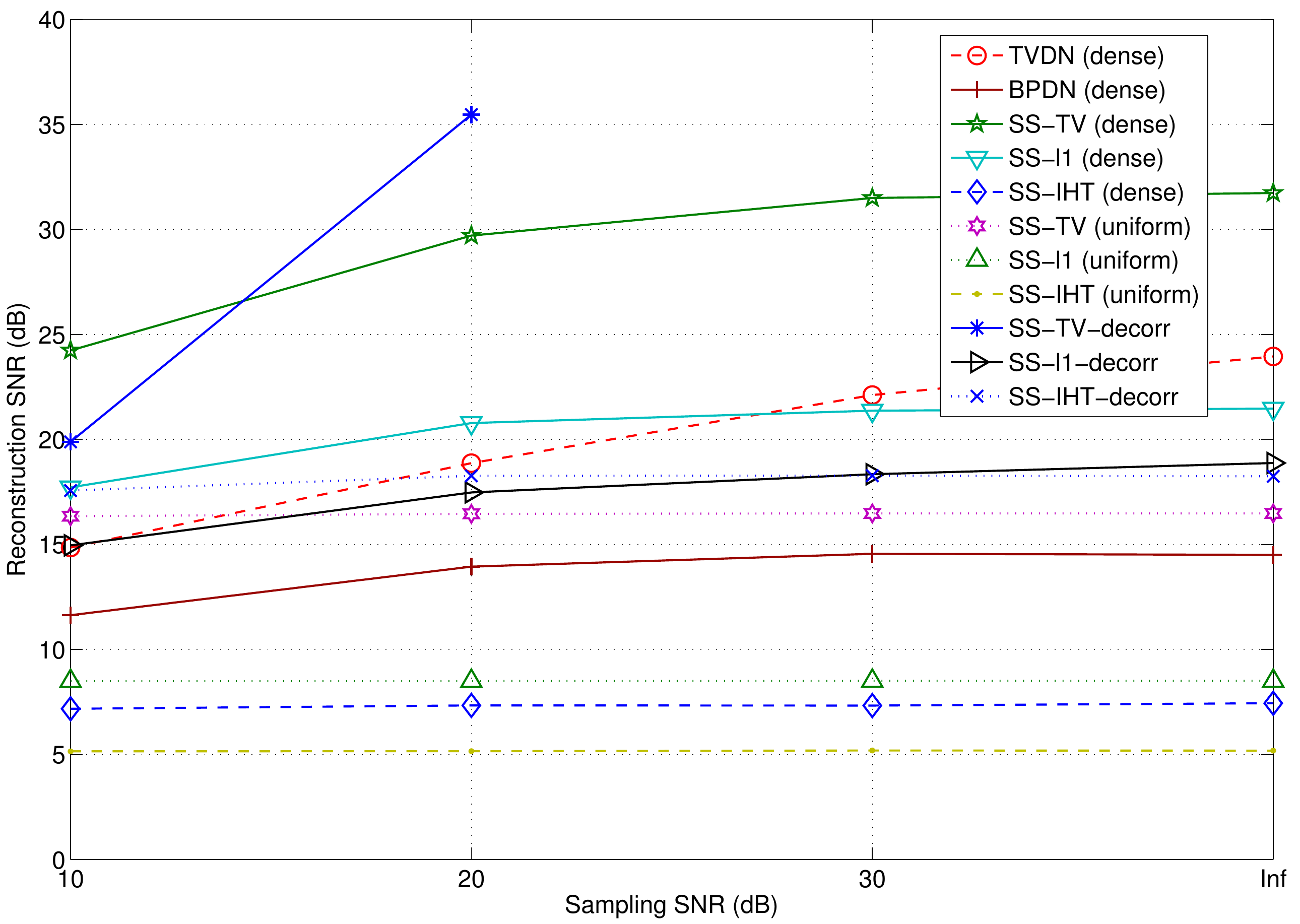}
    \end{minipage}}%
      \caption{Geneva HSI reconstruction performance for different sampling mechanisms and recovery methods. Points with $\infty$ reconstruction SNR (exact recovery) are not plotted. \label{fig:GVA rec}}
\end{figure}
\fi


\subsubsection{Comparison with Classical CS Methods}

We observed that  \fastSSTVfinalProj\ always obtained significantly better results than the classical CS methods in all regimes.

\subsubsection{Source Reconstruction}

We reported in Tab. \ref{tab:SS_performance} the source separation performance of the \SSMethods. Since source images are disjoint, the quality was measured by the source recovery \emph{accuracy}  indicating the percentage of correctly classified pixels in the spatial domain. The method \fastSSTVfinalProj,  based on TV regularization and decorrelation, 
which achieved the best performance for HSI reconstruction also obtain the best performance for source separation.
Figure \ref{fig:rec_sources_Geneva} illustrates the reconstructed sources of different \SSMethods\ 
for various sampling schemes (dense, uniform, decorrelating).
\if@twocolumn
\begin{figure}[t!]
\centering 
 \begin{minipage}[c]{0.46 \textwidth}
  \centering \subfigure[True sources \label{fig:src_Geneva}]{
    \label{}
    \begin{minipage}[b]{ \linewidth}
      \centering \includegraphics[width= \myfigwidth]{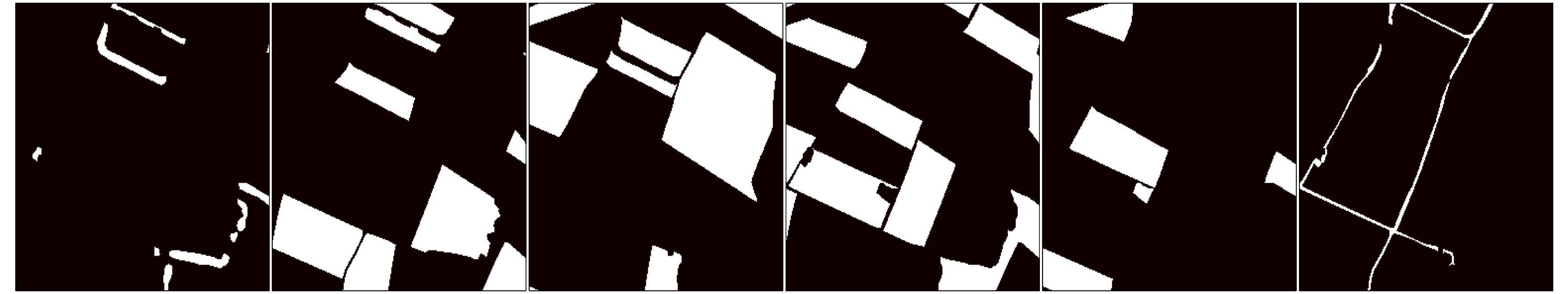}
    \end{minipage}}   
 \centering \subfigure[\fastSSTVfinal\  \emph{(dense sampling)} 
 ]{
    \label{}
    \begin{minipage}[b]{ \linewidth}
      \centering \includegraphics[width= \myfigwidth]{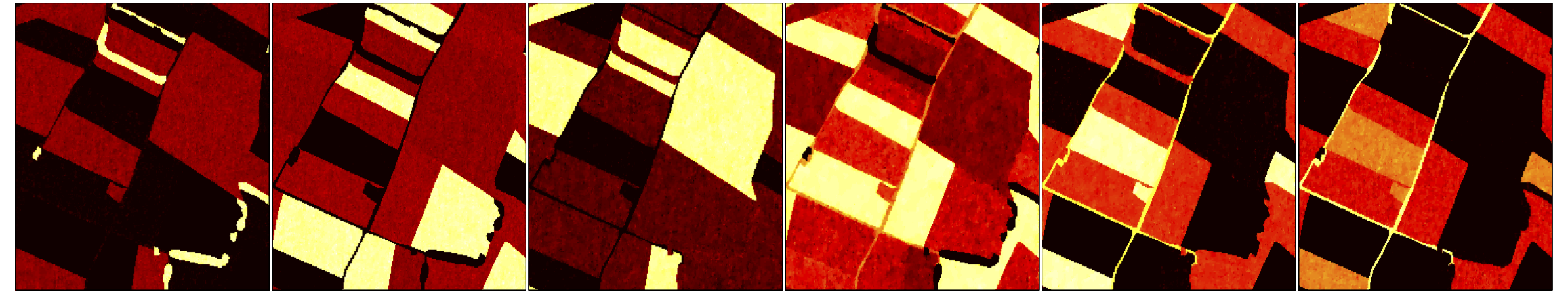}
    \end{minipage}}
      \centering \subfigure[\fastSSTVfinal\  \emph{(uniform non-decorrelating sampling)}
      ]{
    \label{}
    \begin{minipage}[b]{\linewidth}
      \centering \includegraphics[width= \myfigwidth]{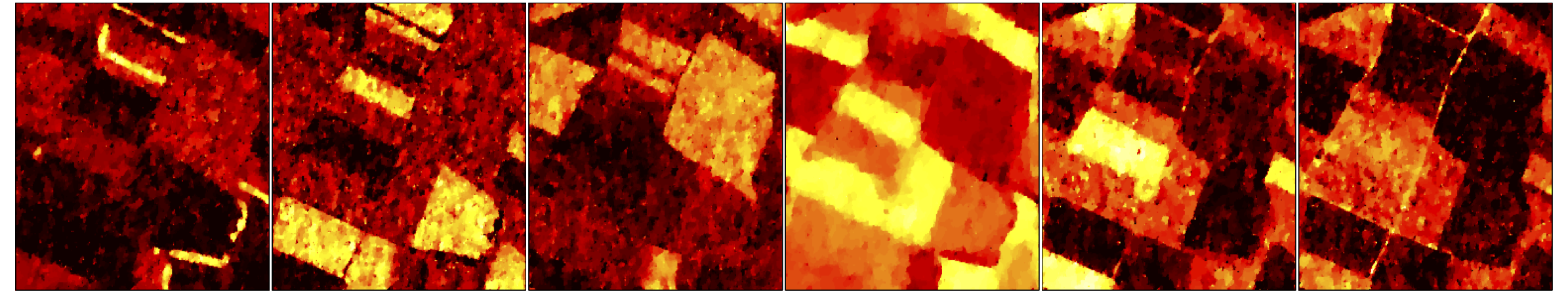}
    \end{minipage}}
     \centering \subfigure[\fastSSTVfinalProj\   \emph{(uniform decorrelating sampling)}
     ]{
    \label{}
    \begin{minipage}[b]{\linewidth}
      \centering \includegraphics[width= \myfigwidth]{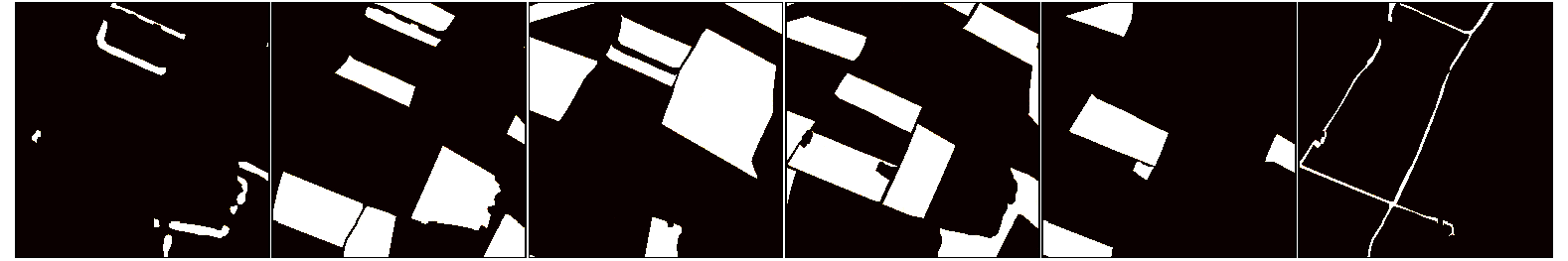}
    \end{minipage}}%
    
    
              \centering \subfigure[\fastSSstIHTfinalProj\  \emph{(uniform decorrelating sampling)}
      ]{
    \label{}
    \begin{minipage}[b]{\linewidth}
      \centering \includegraphics[width= \myfigwidth]{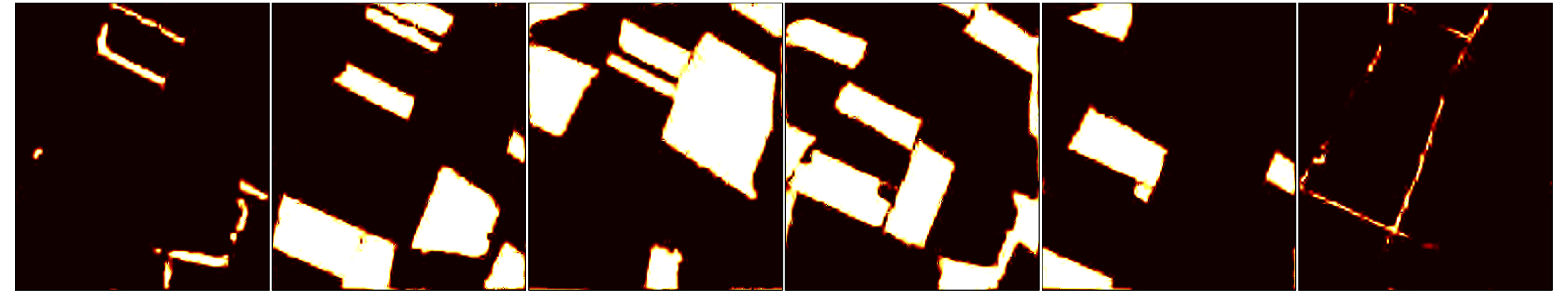}
    \end{minipage}}
    
  \caption{Estimated source images of Geneva HSI for different sampling schemes and recovery methods (subsampling ratio: 1/16, noiseless sampling). \label{fig:rec_sources_Geneva}}
 \end{minipage} \hfill
 \end{figure}

 \fi


\if@twocolumn
\begin{figure*}[t]
  \centering \subfigure[Reference: Sources estimated with a BSS algorithm.]{
    \label{fig:URBAN-GT}
    \begin{minipage}[b]{\linewidth}
      \centering \includegraphics[width= .95\textwidth]{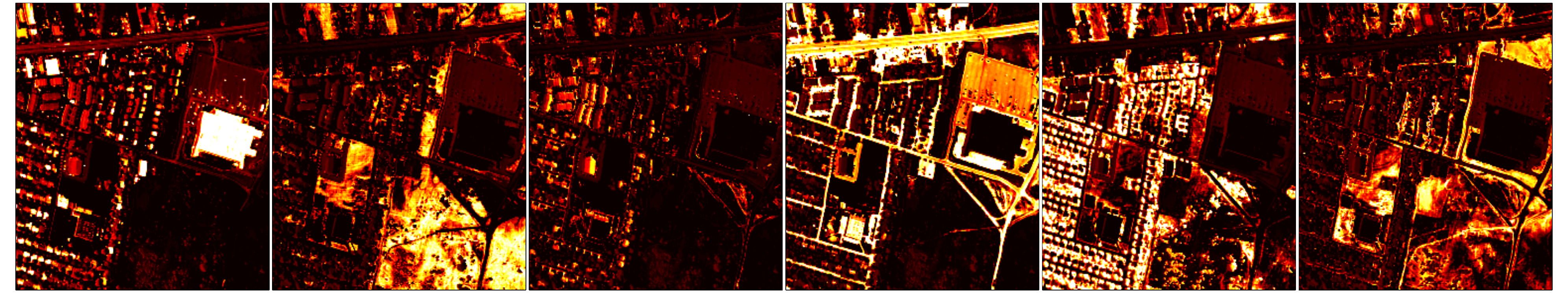}
    \end{minipage}}%
    \hfil
      \centering \subfigure[\fastSSTVfinal\ \emph{(dense sampling)}, source reconstruction SNR: 6.34 dB]{
    \label{fig:URBAN-SS-TV-dense}
    \begin{minipage}[b]{\linewidth}
      \centering \includegraphics[width= .95\textwidth]{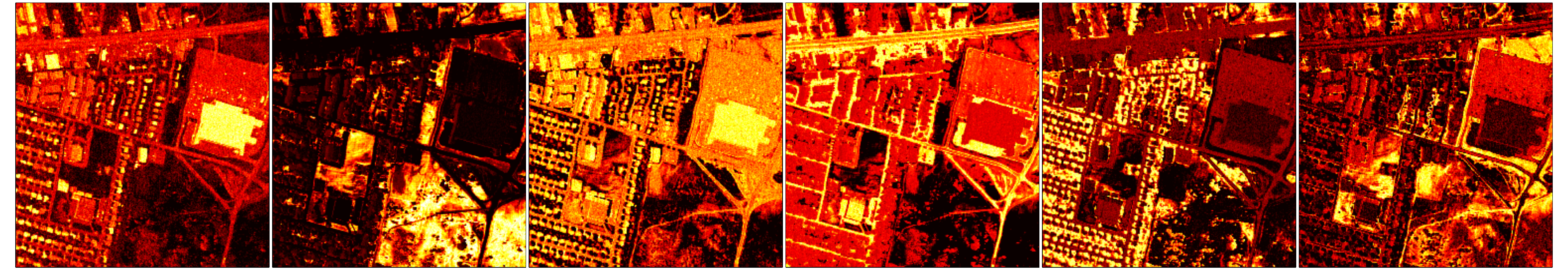}
    \end{minipage}}%
     \hfil
      \centering \subfigure[\fastSSlonefinal\ \emph{(dense sampling)}, source reconstruction SNR: 6.29 dB]{
    \label{}
    \begin{minipage}[b]{\linewidth}
      \centering \includegraphics[width= .95\textwidth]{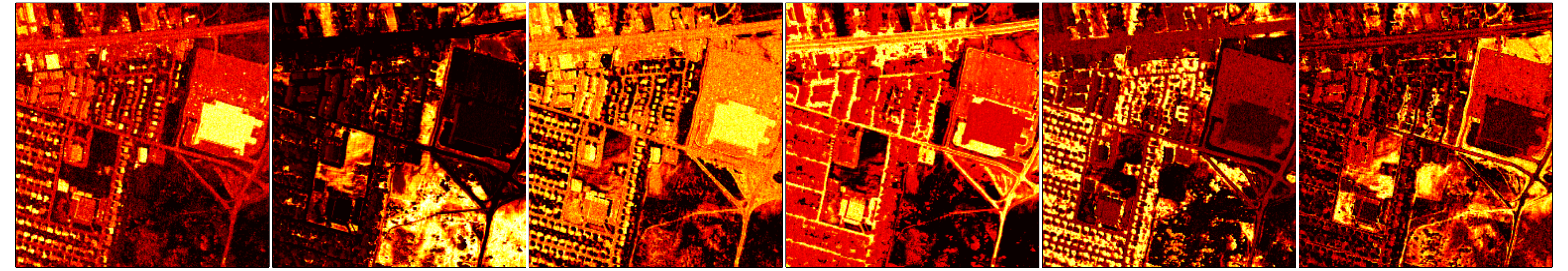}
    \end{minipage}}

   \hfil
   \centering \subfigure[\fastSSTVfinal\ \emph{(uniform non-decorrelating sampling)}, source reconstruction SNR: 1.88 dB]{
  \label{}
    \begin{minipage}[b]{\linewidth}
      \centering \includegraphics[width= .95\textwidth]{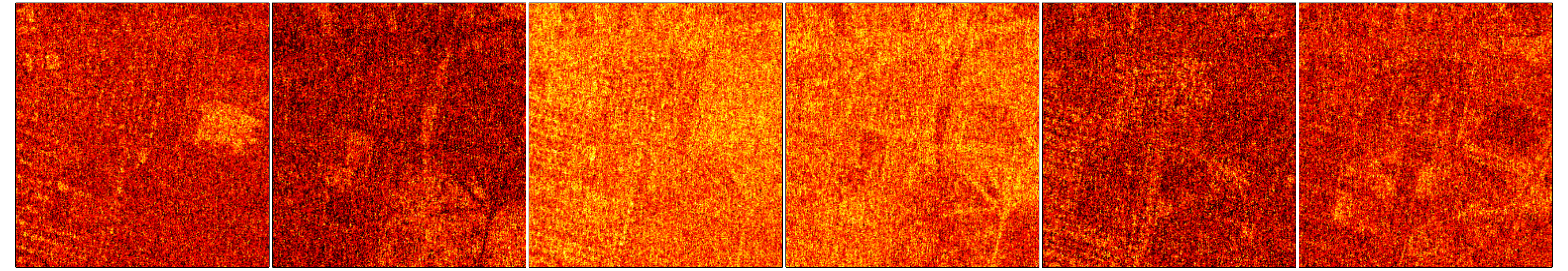}
    \end{minipage}}%
    \hfil
      \centering \subfigure[\fastSSTVfinalProj\ \emph{(uniform decorrelating sampling)}, source reconstruction SNR: 8.64 dB]{
    \label{fig:URBAN-SS-TV-decorr}
    \begin{minipage}[b]{\linewidth}
      \centering \includegraphics[width= .95\textwidth]{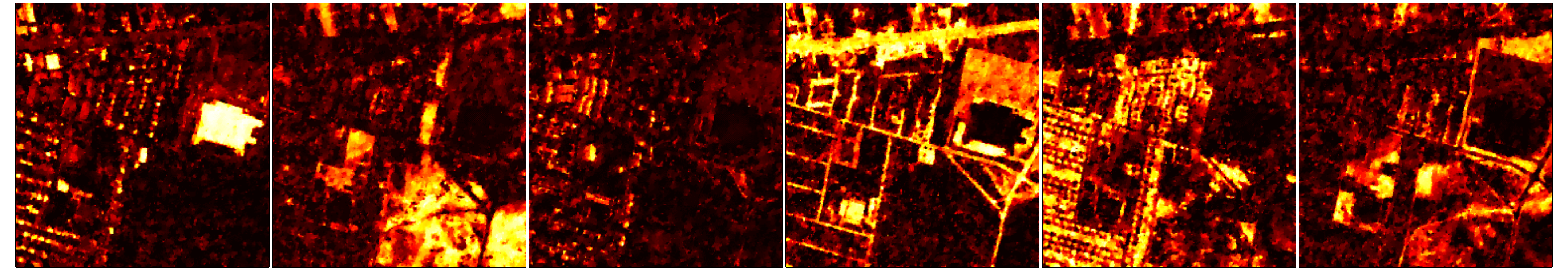}
    \end{minipage}}%
    \hfil
      \centering \subfigure[\fastSSlonefinalProj\ \emph{(uniform decorrelating sampling)}, source reconstruction SNR: 5.65 dB]{
    \label{}
    \begin{minipage}[b]{\linewidth}
      \centering \includegraphics[width= .95\textwidth]{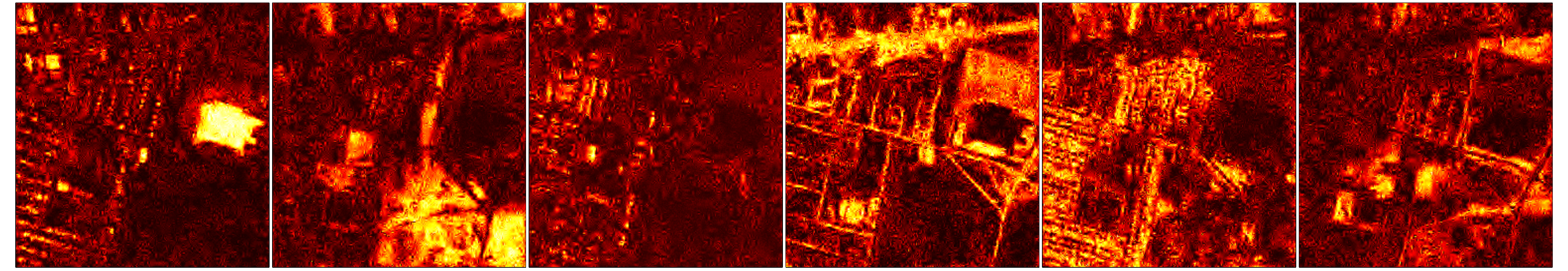}
    \end{minipage}}
       
 \caption{Estimated source images of Urban HSI using different recovery methods (\ie, TV or wavelet $\ell_{1}$ minimization), and for different sampling mechanisms (subsampling ratio: 1/8, noiseless sampling). \label{fig:source recov-URBAN} }
\end{figure*}

\else

\begin{figure*}[h!]
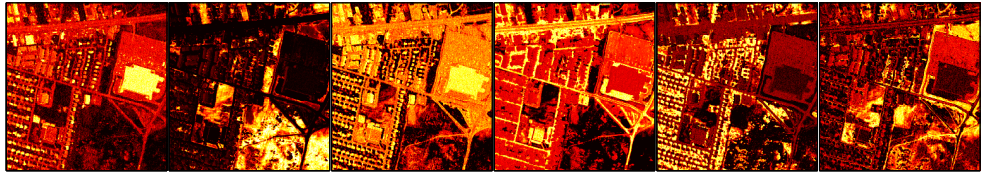
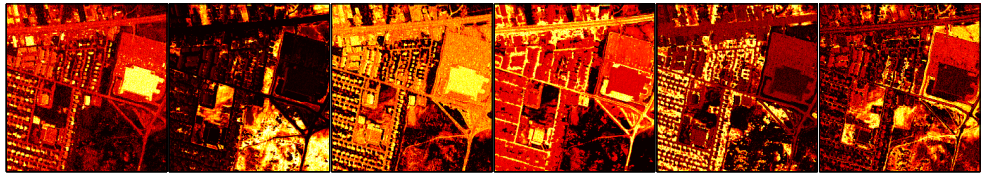
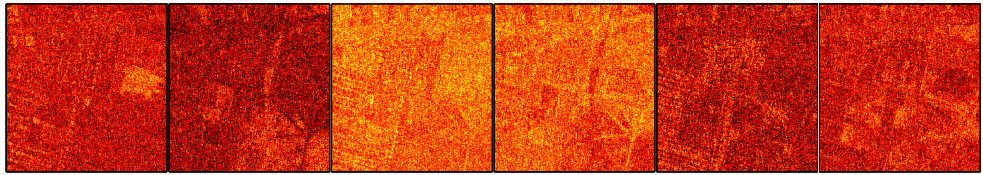
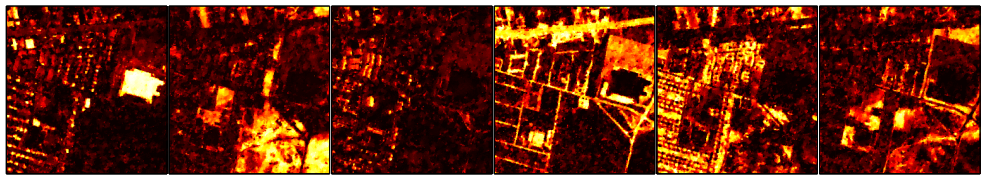
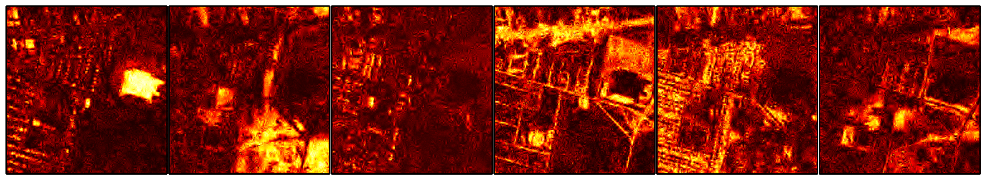

  \centering \subfigure[Reference: Sources estimated with a BSS algorithm.]{
    \label{fig:URBAN-GT}
    \begin{minipage}[b]{\linewidth}
      \centering \includegraphics[width= .8\textwidth]{figures/Urban_GT}
    \end{minipage}}%
    \hfil
      \centering \subfigure[\fastSSTVfinal\ \emph{(dense sampling)}, source reconstruction SNR: 6.34 dB]{
    \label{fig:URBAN-SS-TV-dense}
    \begin{minipage}[b]{\linewidth}
      \centering \includegraphics[width= .8\textwidth]{figures/Urban_Dense_fastSS_TV_final_M_8_SNR_Inf}
    \end{minipage}}%
     \hfil
      \centering \subfigure[\fastSSlonefinal\ \emph{(dense sampling)}, source reconstruction SNR: 6.29 dB]{
    \label{}
    \begin{minipage}[b]{\linewidth}
      \centering \includegraphics[width= .8\textwidth]{figures/Urban_Dense_fastSS_l1_final_M_8_SNR_Inf}
    \end{minipage}}

   \hfil
   \centering \subfigure[\fastSSTVfinal\ \emph{(uniform non-decorrelating sampling)}, source reconstruction SNR: 1.88 dB]{
  \label{}
    \begin{minipage}[b]{\linewidth}
      \centering \includegraphics[width= .8\textwidth]{figures/Urban_Block_diag_fastSS_TV_final_M_8_SNR_Inf}
    \end{minipage}}%
    \hfil
      \centering \subfigure[\fastSSTVfinalProj\ \emph{(uniform decorrelating sampling)}, source reconstruction SNR: 8.64 dB]{
    \label{fig:URBAN-SS-TV-decorr}
    \begin{minipage}[b]{\linewidth}
      \centering \includegraphics[width= .8\textwidth]{figures/Urban_Block_diag_fastSS_TV_final_Proj_M_8_SNR_Inf}
    \end{minipage}}%
    \hfil
      \centering \subfigure[\fastSSlonefinalProj\ \emph{(uniform decorrelating sampling)}, source reconstruction SNR: 5.65 dB]{
    \label{}
    \begin{minipage}[b]{\linewidth}
      \centering \includegraphics[width= .8\textwidth]{figures/Urban_Block_diag_fastSS_l1_final_Proj_M_8_SNR_Inf}
    \end{minipage}}
       
 \caption{Estimated source images of Urban HSI using different recovery methods (\ie, TV or wavelet $\ell_{1}$ minimization), and for different sampling mechanisms (subsampling ratio: 1/8, noiseless sampling). \label{fig:source recov-URBAN} }
\end{figure*}

\fi

 \if@twocolumn
 \else
 \begin{figure}[t]
\centering
\includegraphics[width= .85\textwidth]{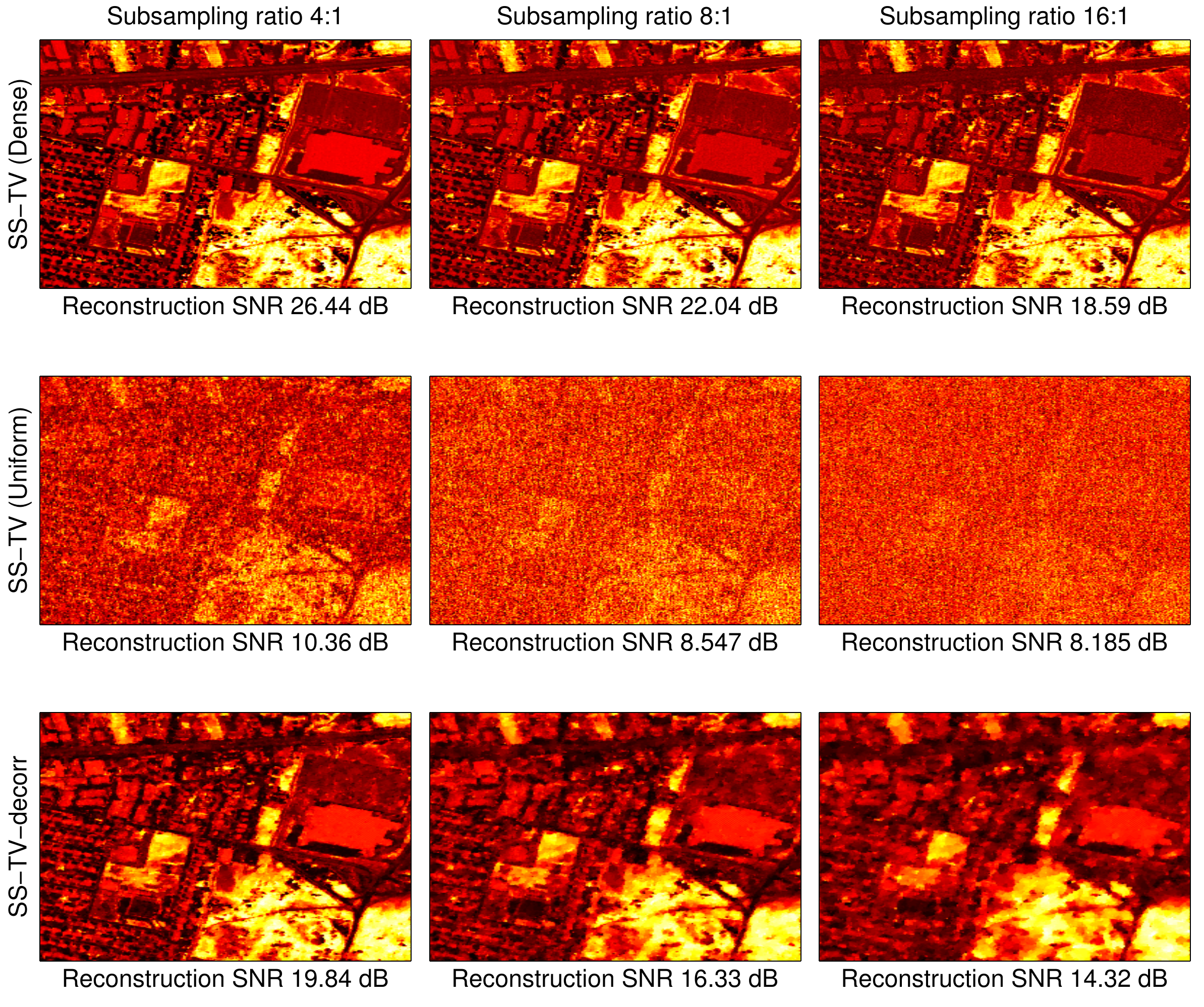}
         \caption{Reconstructed Urban HSI at spectral band 33, using SS methods based on TV minimization, for various sampling mechanisms (Dense, Uniform non-decorrelating, Uniform decorrelating) and subsampling ratios. \label{fig:URBAN-rec} }
\end{figure}
\fi

%



\subsection{The Urban HSI}

In order to evaluate different approaches on a real HSI, we consider the Urban HSI of size $256 \times 256 \times 171$ 
which was obtained from the site \cite{usarmyTEC} of the US Army Topographic Engineering Center.

As the ground truth of this image (\ie, the true source images and their corresponding spectral signatures) is not available,  we first separate the underlying sources using a \emph{blind} source separation algorithm for fully-sampled HSI \cite{arberet2010hyper} and later, use these separated sources, depicted in Fig. \ref{fig:URBAN-GT}, as a reference.
 Figure \ref{fig:source recov-URBAN} demonstrates the reconstructed sources of Urban using our proposed SS approaches based on convex minimization, for different noiseless sampling mechanisms (dense, uniform, uniform-decorrelating) and for a fixed subsampling ratio.\footnote{As the source images of Urban are not spatially disjoint, we do not apply  
 Algo. 2.}
 Moreover, Figure \ref{fig:URBAN-rec} shows the reconstructed Urban HSI for a certain spectral band, using the source images estimated by the SS methods based on TV minimization (\ie, $\fastSSTVfinal$ and $\fastSSTVfinalProj$).




\if@twocolumn
\begin{figure}[ht]
\centering
\includegraphics[width= .49\textwidth]{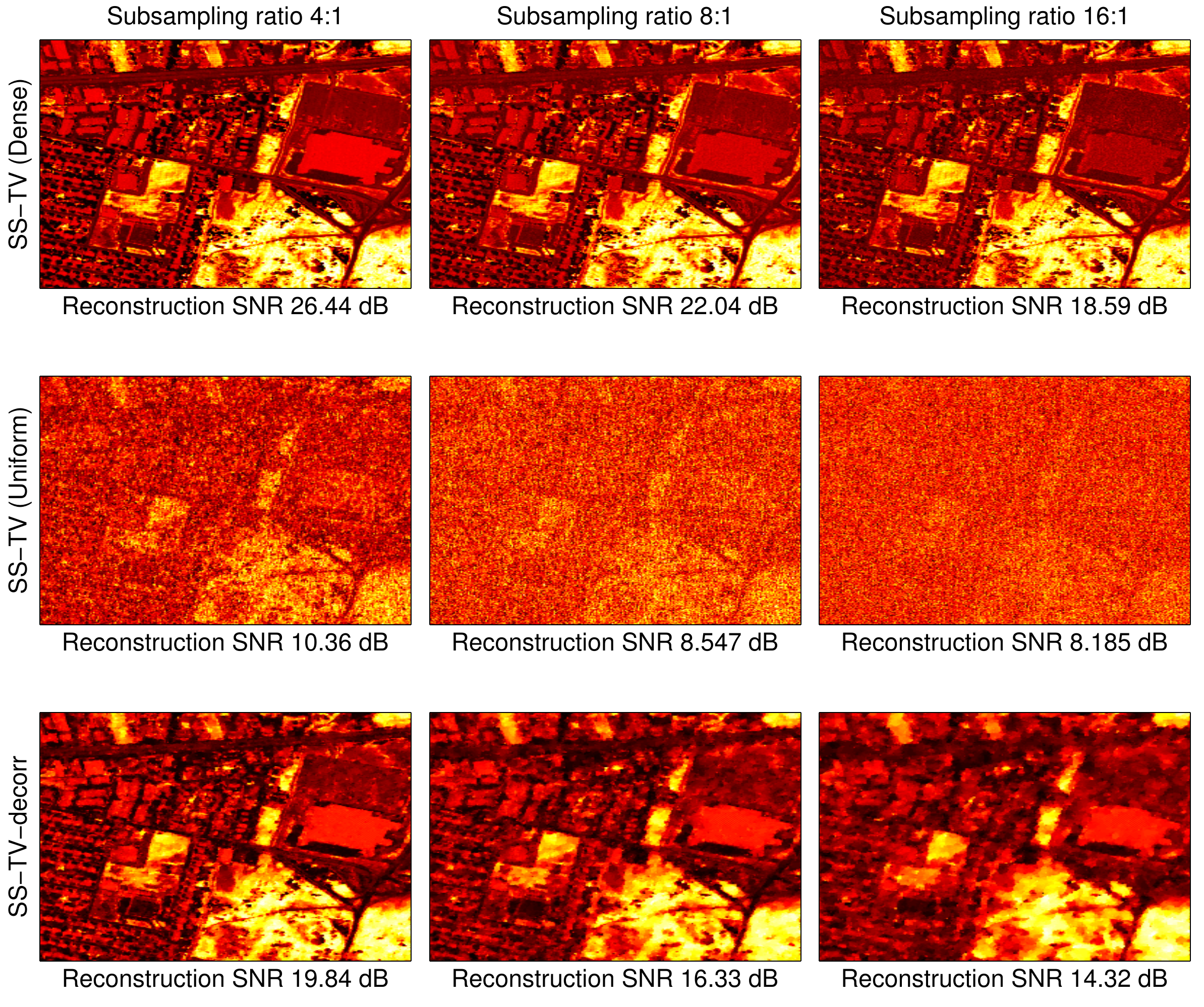}
         \caption{Reconstructed Urban HSI at spectral band 33, using SS methods based on TV minimization, for various sampling mechanisms (Dense, Uniform non-decorrelating, Uniform decorrelating) and subsampling ratios. \label{fig:URBAN-rec} }
\end{figure}
\fi

\paragraph{Results}
Similar to our previous experiment, we observe that for a uniform (non-decorrelating) sampling scheme \fastSSTVfinal\ has very poor recovery performance. Meanwhile, adding a decorrelation step results in a significant improvement in source recovery.
As we can see in Figure \ref{fig:URBAN-SS-TV-dense}, the estimated source images using \fastSSTVfinal\ for a dense sampling scheme have better spatial resolutions, 
but are not as well separated as with the \fastSSTVfinalProj\ method.


\paragraph{Computational Performance}
Decorrelation step massively decreases the computational complexity. \fastSSTVfinalProj\ performs within 20 minutes whereas \fastSSTVfinal\ for a dense sampling scheme requires more than 80 hours of computations! 
The classical \solveBPDNone\ and \solveTVDN\ methods take between $3$ to $14$ hours, as the corresponding $\ell_1$ or TV minimization runs over a large number of channels (rather than few underlying sources). We ran all the codes on a Mac Pro 2.26  GHz Intel CPU, 16 GB RAM computer.

\subsection{Conclusion on the Experiments}
The decorrelation step is of great benefit, and the proposed method \fastSSTVfinalProj, based on TV regularization and decorrelation, outperforms significantly all the other methods for HSI reconstruction and source estimation for all tested SNRs and  sampling rates.
Moreover \fastSSTVfinalProj\ is clearly the fastest method  and is more than $40$ times faster than the classical \solveTVDN.

While finalizing this work we became aware of a recent paper \cite{CS-HSI-UNMIX} that proposes a source recovery approach similar to \eqref{HSI Analysis SS}, albeit for the particular case of uniform sampling and TV regularization. The authors also use a "SVD preprocessing" step for dimensionality reduction and denoising that, contrary to our decorrelation step, does not cancel the effects of the conditioning of the mixing matrix. 
Comparing the fourth source image in Figure \ref{fig:URBAN-SS-TV-decorr}, corresponding to the ``roads'', with the similar source recovered in Figure 6 in \cite{CS-HSI-UNMIX} indicates that \fastSSTVfinalProj\  achieves similar or better separation performance with only half the measurements rate. Additionally we provide a theoretical analysis for the compressive source separation problem, considering various sampling schemes and multiple  recovery methods.


\section{Conclusion}

In this paper, we exploited a linear mixture of sources model into a Compressed Sensing (CS) scheme for
multichannel signal acquisition and source separation with a particular focus on hyperspectral images (HSI). 
We study three different acquisition schemes (dense, uniform and decorrelated) theoretically and experimentally,
and showed that the decorrelating scheme enhances drastically the recovery of the spectral data and its sources.
Indeed, our theoretical analysis showed that, using this scheme, and contrary to the traditional CS approach, 
the number of measurements does not scale with the number of channels and does not depend on the conditioning of the mixing matrix, as long as the mixed spectra are linearly independent. This leads to a strong reduction in the number of needed measurements for a given reconstruction error.
We also provided algorithms that reconstruct the multichannel signal (more particularly HSI)  and its sources, by exploiting both sparsity of the signal at each channel and the correlation of the signals along the channels.
We provided experiments on HSI and showed that we can reconstruct both the HSI and its sources with far fewer measurements and less computational effort than traditional CS approaches. Finally, we showed that it is possible to accurately recover the sources directly from the compressed measurements, avoiding to run a source separation algorithm on the high-dimensional raw data.

Extension of this work includes dealing with non-linear mixture of sources as well as dealing with the difficult problem of recovering the sources and the mixing system from the compressed measurements.

\section{Appendix}
\appendices

\subsection{Proof of Theorem \ref{thSS}}\label{sec:ch5-proof}

Let $\theta\in \Rbb^{d}$ be the original vector we aim to recover from its CS measurements $y$ with $y = A\D\theta + z$ and $\|z\|_{2}\leq \varepsilon$, and let $\widehat \theta$  be the solution of the $\ell_{1}$ minimization \eqref{dic CS}. The reconstruction error is denoted $h=\widehat \theta -\theta$. Let $\tcal_{0} \subseteq \{0,\ldots,d\}$ be the set that contains the indices of the $k$ coefficients of $\theta$ having the largest magnitudes and, $\tcal_{0}^{c}$ the complement set of $\tcal_{0}$. Let $\theta_{\tcal}$ denote a vector of the same size as $\theta$ whose elements indexed by the set $\tcal$ are identical to that of $\theta$ and zero elsewhere.

Minimizing the $\ell_{1}$ norm in \eqref{dic CS} implies
\begin{eqnarray}
\|\theta\|_{1} &\geq& \|\theta + h\|_{1} \nonumber \\
&=& \|\theta_{\tcal_{0}} + h_{\tcal_{0}}\|_{1} + \|\theta_{\tcal^{c}_{0}} + h_{\tcal^{c}_{0}}\|_{1} \nonumber \\
&\geq& \|\theta_{\tcal_{0}} \|_{1}- \| h_{\tcal_{0}} \|_{1} - \|\theta_{\tcal^{c}_{0}} \|_{1} + \| h_{\tcal^{c}_{0}} \|_{1} , \nonumber
\end{eqnarray}
and therefore,
\begin{eqnarray}\label{ch5-ineq1}
\| h_{\tcal^{c}_{0}} \|_{1} \leq \| h_{\tcal_{0}} \|_{1} + 2 \| \theta_{\tcal^{c}_{0}} \|_{1}. 
\end{eqnarray}
Let $\tcal_{1}$ be the set that contains the indices of the $\tau k$ coefficients of $\theta_{\tcal^{c}_{0}} $ having the largest magnitudes, $\tcal_{2}$ the set containing the indices of the second $\tau k$ largest coefficients of  $\theta_{\tcal^{c}_{0}} $, and so on. With this decomposition, $\forall j \geq 2$ we have:
\[
\|h_{\tcal_{j}}\|_{2} \leq (\tau k )^{-1/2}\,\|h_{\tcal_{j-1}}\|_{1},
\] 
and thus,
\[
\sum_{j\geq 2} \|h_{\tcal_{j}}\|_{2}  \leq (\tau k )^{-1/2}\,\|h_{\tcal^{c}_{0}}\|_{1}.
\]
Now according to \eqref{ch5-ineq1} and since $h_{\tcal_{0}} $ is $k$-sparse we have  
\begin{eqnarray}\label{ch5-ineq2}
\sum_{j\geq 2} \|h_{\tcal_{j}}\|_{2}  \leq \tau^{-1/2} \| h_{\tcal_{0}} \|_{2}+ 2(\tau k)^{-1/2}\| \theta_{\tcal^{c}_{0}} \|_{1}.
\end{eqnarray}
On the other hand, since both $\theta$ and $\widehat \theta$ satisfy the fidelity constraint of \eqref{dic CS}, we have
\begin{eqnarray}
\|A\D h\|_{2} \leq \|y - A\D \theta\|_{2} + \|y - A\D \widehat\theta\|_{2} \leq 2\varepsilon. \nonumber
\end{eqnarray}
Let's define $\tcal_{01}:=\tcal_{0} \cup \tcal_{1}$ and $\gamma = \tau+1$. According to the last inequality we can write
\if@twocolumn 
\begin{eqnarray}
2\varepsilon &\geq& \|A\D h\|_{2} \nonumber \\
&\geq& \|A\D h_{\tcal_{01}}\|_{2} - \sum_{j\geq 2 } \|A\D h_{\tcal_{j}}\|_{2} \nonumber\\
&\geq& \sqrt{1-\delta^{*}_{\gamma k} } \, \|\D h_{\tcal_{01}}\|_{2} -  \sqrt{1+\delta^{*}_{\tau k} } \, \sum_{j\geq 2 } \|\D h_{\tcal_{j}}\|_{2}\nonumber \\
&\geq& \lo_{\gamma k}(\D)  \sqrt{1-\delta^{*}_{\gamma k} } \,\| h_{\tcal_{01}}\|_{2} \nonumber \\
&& \quad - \  \up_{\tau k}(\D) \sqrt{1+\delta^{*}_{\tau k} }\, \sum_{j\geq 2 } \|h_{\tcal_{j}}\|_{2} \nonumber \\
&\geq& \lo_{\gamma k}(\D)  \sqrt{1-\delta^{*}_{\gamma k} } \| h_{\tcal_{01}}\|_{2}  \nonumber \\
&& - \  \up_{\tau k}(\D) \sqrt{1+\delta^{*}_{\tau k} }\Big( \tau^{-1/2} \| h_{\tcal_{0}} \|_{2}\nonumber \\
&& \qquad \qquad  \qquad \qquad \qquad + 2(\tau k)^{-1/2}\| \theta_{\tcal^{c}_{0}} \|_{1}\Big).\nonumber
\end{eqnarray}
\else 
\begin{eqnarray}
2\varepsilon &\geq& \|A\D h\|_{2} \nonumber \\
&\geq& \|A\D h_{\tcal_{01}}\|_{2} - \sum_{j\geq 2 } \|A\D h_{\tcal_{j}}\|_{2} \nonumber\\
&\geq& \sqrt{1-\delta^{*}_{\gamma k} } \, \|\D h_{\tcal_{01}}\|_{2} -  \sqrt{1+\delta^{*}_{\tau k} } \, \sum_{j\geq 2 } \|\D h_{\tcal_{j}}\|_{2}\nonumber \\
&\geq& \lo_{\gamma k}(\D)  \sqrt{1-\delta^{*}_{\gamma k} } \,\| h_{\tcal_{01}}\|_{2}
 - \  \up_{\tau k}(\D) \sqrt{1+\delta^{*}_{\tau k} }\, \sum_{j\geq 2 } \|h_{\tcal_{j}}\|_{2} \nonumber \\
&\geq& \lo_{\gamma k}(\D)  \sqrt{1-\delta^{*}_{\gamma k} } \| h_{\tcal_{01}}\|_{2} 
 - \  \up_{\tau k}(\D) \sqrt{1+\delta^{*}_{\tau k} }\Big( \tau^{-1/2} \| h_{\tcal_{0}} \|_{2}
 + 2(\tau k)^{-1/2}\| \theta_{\tcal^{c}_{0}} \|_{1}\Big).\nonumber
\end{eqnarray}
\fi
The third inequality follows from definition of the D-RIP (see Definition \ref{ch5-def2}) which holds for the matrix $A$, together with the fact that $h_{\tcal_{01}}$ and $h_{\tcal_{j}}$ ($\forall j\geq 2$) are respectively $\gamma k$ and $\tau k$ sparse. The fourth inequality follows from the definition of the A-RIP that holds for matrix $\D$ (see Definition \ref{ch5-def1}), and finally the last inequality uses \eqref{ch5-ineq2}. We apply the bounds $\delta^{*}_{\tau k}\leq \delta^{*}_{\gamma k} $, $\up_{\tau k}(\D)\leq \up_{\gamma k}(\D)$ and $ \| h_{\tcal_{0}} \|_{2} \leq  \| h_{\tcal_{01}} \|_{2}$ in the last inequality and we deduce the following bound: 
\begin{eqnarray}\label{ch5-ineq3}
\| h_{\tcal_{01}}\|_{2} \leq \alpha k^{-1/2}\| \theta_{\tcal^{c}_{0}} \|_{1} + \beta \varepsilon,
\end{eqnarray}
where the constants $\alpha, \beta$ are
$\alpha = \frac{2}  { \xi_{\gamma k}^{-1}(\D) \sqrt{\tau \left( \frac{1-\delta^{*}_{\gamma k}} {1+\delta^{*}_{\gamma k}} \right)} - 1  },$ \quad and \quad 
$\beta = \frac{2\, \up_{\gamma k}(\D) \sqrt{\tau(1+\delta^{*}_{\gamma k})} }  { \xi_{\gamma k}^{-1}(\D) \sqrt{\tau \left( \frac{1-\delta^{*}_{\gamma k}} {1+\delta^{*}_{\gamma k}} \right)} - 1  }.$
\newline
Now if we set $\tau \geq 2\xi_{\gamma k}^{2}(\D)$ (equivalently, $\gamma \geq 1+2\xi_{\gamma k}^{2}(\D)$), it is sufficient to have $\delta^{*}_{\gamma k} < 1/3$ so that $\alpha$ and $\beta$ remain positive.
Finally we conclude the proof of Theorem \ref{thSS} by using the inequalities \eqref{ch5-ineq2} and \eqref{ch5-ineq3} to bound the whole error term as follows:
\begin{eqnarray}
\|h\|_{2} &\leq& \|h_{\tcal_{01}}\|_{2} + \sum_{j\geq 2} \|h_{\tcal_{j}}\|_{2}  \nonumber \\
&\leq&  (1+\tau^{-1/2}) \| h_{\tcal_{01}} \|_{2}+ 2(\tau k)^{-1/2}\| \theta_{\tcal^{c}_{0}} \|_{1} \nonumber\\
&\leq& c'_{0} k^{-1/2} \| \theta_{\tcal^{c}_{0}} \|_{1}+c'_{1}\varepsilon, \nonumber 
\end{eqnarray}
where, the constants of the error bound are $c'_{0} = \alpha + (2+\alpha)\tau^{-1/2}$ and $c'_{1} = \beta(1+\tau^{-1/2})$.

%
%
%
%

\bibliographystyle{IEEEtran}
\bibliography{Ch4}

\begin{thebibliography}{10}
\providecommand{\url}[1]{#1}
\csname url@samestyle\endcsname
\providecommand{\newblock}{\relax}
\providecommand{\bibinfo}[2]{#2}
\providecommand{\BIBentrySTDinterwordspacing}{\spaceskip=0pt\relax}
\providecommand{\BIBentryALTinterwordstretchfactor}{4}
\providecommand{\BIBentryALTinterwordspacing}{\spaceskip=\fontdimen2\font plus
\BIBentryALTinterwordstretchfactor\fontdimen3\font minus
  \fontdimen4\font\relax}
\providecommand{\BIBforeignlanguage}[2]{{%
\expandafter\ifx\csname l@#1\endcsname\relax
\typeout{** WARNING: IEEEtran.bst: No hyphenation pattern has been}%
\typeout{** loaded for the language `#1'. Using the pattern for}%
\typeout{** the default language instead.}%
\else
\language=\csname l@#1\endcsname
\fi
#2}}
\providecommand{\BIBdecl}{\relax}
\BIBdecl

\bibitem{donoho2006compressed}
D.~Donoho, ``{Compressed sensing},'' \emph{IEEE Trans. on Inf. Theory},
  vol.~52, no.~4, pp. 1289--1306, 2006.

\bibitem{CRT-stable2005}
E.~J. Candes, J.~Romberg, and T.~Tao, ``Stable signal recovery from incomplete
  and inaccurate measurements.'' \emph{Pure Appl. Math.}, vol.~59, pp.
  1207--1223, 2005.

\bibitem{duarte-kronecker}
M.~Duarte and R.~Baraniuk, ``{Kronecker Compressive Sensing},'' \emph{to appear
  in the IEEE Trans. on Image Processing}, 2009.

\bibitem{SinglepixelHSI}
T.~Sun and K.~Kelly, ``Compressive sensing hyperspectral imager,'' \emph{Comp.
  Optical Sensing and Imaging (COSI), San Jose, CA, Oct. 2009.}

\bibitem{CASSI}
A.~Wagadarikar, R.~John, R.~Willett, and D.~Brady, ``Single disperser design
  for coded aperture snapshot spectral imaging,'' \emph{Applied Optics},
  vol.~47, pp. B44--B51, 2008.

\bibitem{keshava2002spectral}
N.~Keshava and J.~Mustard, ``Spectral unmixing,'' \emph{Signal Processing
  Magazine, IEEE}, vol.~19, no.~1, pp. 44--57, 2002.

\bibitem{nascimento2005vertex}
J.~Nascimento and J.~Dias, ``Vertex component analysis: A fast algorithm to
  unmix hyperspectral data,'' \emph{Geoscience and Remote Sensing, IEEE
  Transactions on}, vol.~43, no.~4, pp. 898--910, 2005.

\bibitem{1677768}
J.~Wang and C.-I. Chang, ``Applications of independent component analysis in
  endmember extraction and abundance quantification for hyperspectral
  imagery,'' \emph{Geoscience and Remote Sensing, IEEE Transactions on},
  vol.~44, no.~9, sept. 2006.

\bibitem{1261124}
H.~Ren and C.-I. Chang, ``Automatic spectral target recognition in
  hyperspectral imagery,'' \emph{Aerospace and Electronic Systems, IEEE
  Transactions on}, vol.~39, no.~4, pp. 1232 -- 1249, oct. 2003.

\bibitem{arberet2010hyper}
S.~Arberet, ``Hyper-demix: Blind source separation of hyperspectral images
  using local ml estimates,'' in \emph{Image Processing (ICIP), 2010 17th IEEE
  International Conference on}.\hskip 1em plus 0.5em minus 0.4em\relax IEEE,
  2010, pp. 1393--1396.

\bibitem{golbabaee2010CSBSS}
M.~Golbabaee, S.~Arberet, and P.~Vandergheynst, ``{Multichannel compressed
  sensing via source separation for hyperspectral images},'' in \emph{Eusipco},
  2010.

\bibitem{golbabaee2010distributed}
------, ``Distributed compressed sensing of hyperspectral images via blind
  source separation,'' in \emph{Signals, Systems and Computers (ASILOMAR), 2010
  Conference Record of the Forty Fourth Asilomar Conference on}.\hskip 1em plus
  0.5em minus 0.4em\relax IEEE, 2010.

\bibitem{mallat1993matching}
S.~Mallat and Z.~Zhang, ``Matching pursuits with time-frequency dictionaries,''
  \emph{Signal Processing, IEEE Transactions on}, vol.~41, no.~12, pp.
  3397--3415, 1993.

\bibitem{Natarajan_1995}
B.~K. Natarajan, ``Sparse approximate solutions to linear systems,'' \emph{SIAM
  Journal on Computing}, vol.~24, no.~2, p. 227, 1995.

\bibitem{IHT-sparse}
T.~Blumensath and M.~Davies, ``Iterative thresholding for sparse
  approximations,'' \emph{Journal of Fourier Analysis and Applications},
  vol.~14, pp. 629--654, 2008.

\bibitem{candes2008restricted}
E.~Cand{\`e}s, ``The restricted isometry property and its implications for
  compressed sensing,'' \emph{Comptes Rendus Mathematique}, vol. 346, no. 9-10,
  pp. 589--592, 2008.

\bibitem{JL-lemma}
R.~Baraniuk, M.~Davenport, R.~DeVore, and M.~Wakin, ``A simple proof of the
  restricted isometry property for random matrices,'' \emph{Constructive
  Approximation}, vol.~28, pp. 253--263, 2008.

\bibitem{TV-ROF}
\BIBentryALTinterwordspacing
L.~Rudin, S.~Osher, and E.~Fatemi, ``Nonlinear total variation based noise
  removal algorithms,'' \emph{Physica D: Nonlinear Phenomena}, vol.~60, pp. 259
  -- 268, 1992. [Online]. Available:
  \url{http://www.sciencedirect.com/science/article/pii/016727899290242F}
\BIBentrySTDinterwordspacing

\bibitem{DicCS}
H.~Rauhut, K.~Schnass, and P.~Vandergheynst, ``Compressed sensing and redundant
  dictionaries,'' \emph{Information Theory, IEEE Transactions on}, vol.~54,
  no.~5, pp. 2210 --2219, may 2008.

\bibitem{Candes-Eldar}
E.~J. Cand\`es, Y.~C. Eldar, D.~Needell, and P.~Randall, ``Compressed sensing
  with coherent and redundant dictionaries,'' \emph{Applied and Computational
  Harmonic Analysis}, vol.~31, no.~1, pp. 59 -- 73, 2011.

\bibitem{proximal-splitting}
P.~L. Combettes and J.~C. Pesquet, ``Proximal splitting methods in signal
  processing,'' \emph{in: Fixed-Point Algorithms for Inverse Problems in
  Science and Engineering, Springer-Verlag}, vol.~49, pp. 185--212, 2011.

\bibitem{TV-chambolle}
A.~Chambolle, ``An algorithm for total variation minimization and
  applications,'' \emph{Journal of Mathematical Imaging and Vision}, vol.~20,
  pp. 89--97, 2004.

\bibitem{Duchi}
J.~Duchi, S.~Shalev-Shwartz, Y.~Singer, and T.~Chandra, ``Efficient projections
  onto the l1-ball for learning in high dimensions,'' in \emph{Proceedings of
  the 25th international conference on Machine learning}, ser. ICML '08, 2008,
  pp. 272--279.

\bibitem{FadiliS09}
M.~Fadili and J.~Starck, ``Monotone operator splitting for optimization
  problems in sparse recovery,'' in \emph{Image Processing (ICIP), 2009 16th
  IEEE International Conference on}.\hskip 1em plus 0.5em minus 0.4em\relax
  IEEE, 2009, pp. 1461--1464.

\bibitem{RCromberg}
J.~Romberg, ``Compressive sensing by random convolution,'' \emph{SIAM J.
  Imaging Sciences}, 2009.

\bibitem{speclabUSGS}
\BIBentryALTinterwordspacing
 [Online]. Available: \url{http://speclab.cr.usgs.gov/spectral.lib06}
\BIBentrySTDinterwordspacing

\bibitem{usarmyTEC}
\BIBentryALTinterwordspacing
 [Online]. Available: \url{http://www.agc.army.mil/hypercube/}
\BIBentrySTDinterwordspacing

\bibitem{CS-HSI-UNMIX}
C.~Li, T.~Sun, K.~Kelly, and Y.~Zhang, ``A compressive sensing and unmixing
  scheme for hyperspectral data processing,'' \emph{Image Processing, IEEE
  Transactions on}, vol.~21, no.~3, pp. 1200 --1210, march 2012.

\end{thebibliography}

\end{document}